\documentclass[reprint,superscriptaddress,twocolumn,10pt]{revtex4-1}
\usepackage{bbm}
\usepackage{physics}
\usepackage{amsmath}
\usepackage{epsfig}
\usepackage{subfigure,mathrsfs}
\usepackage{array}
\usepackage{amssymb}
\usepackage{braket}
\usepackage{float}
\usepackage{lmodern,amssymb}
\usepackage{physics}
\usepackage[dvipsnames]{xcolor}
\usepackage{lipsum}

\newcommand{\beq}[0]{\begin{equation}}
\newcommand{\eeq}[0]{\end{equation}}

\def\be{\begin{equation}}
\def\ee{\end{equation}}
\def\bea{\begin{eqnarray}}
\def\eea{\end{eqnarray}}
\newcommand{\ba}{\begin{eqnarray}}
\newcommand{\ea}{\end{eqnarray}}
\usepackage{hyperref}

\usepackage[bb=boondox]{mathalfa}
\usepackage{braket}

\begin{document}
\title{Graph theoretic analysis of three-terminal quantum dot thermocouples: Onsager relations and spin-thermoelectric effects}

\author{Nikhil Gupt}
\affiliation{Indian Institute of Technology Kanpur, 
	Kanpur, Uttar Pradesh 208016, India}
\author{Shuvadip Ghosh}
\affiliation{Indian Institute of Technology Kanpur, 
	Kanpur, Uttar Pradesh 208016, India}
\author{Arnab Ghosh}
\thanks{arnab@iitk.ac.in}
\affiliation{Indian Institute of Technology Kanpur, 
	Kanpur, Uttar Pradesh 208016, India}

\begin{abstract}
We introduce a simplified model for a three-terminal quantum thermocouple consisting of two strongly-coupled quantum dots. To elucidate spin-dependent
Seebeck and Peltier effects, we employ a microscopic Hamiltonian and map the Lindblad master
equation onto a quantum transition network, capturing the key working principles for
both reciprocal effects. Our analysis reveals quantum thermodynamic networks encompassing both Coulomb interaction and
spin-flipping processes, lead to the emergence of spin-thermolectric effects. Using algebraic graph theory, we recover the phenomenological law of irreversible thermodynamics from the stochastic version of the entropy production rate expressed in terms of cycle
flux and cycle forces. Remarkably, Onsager reciprocity and Kelvin relation for transport coefficients find their premises in the properties of cycle flux trajectories within the quantum transition network. This underscores the universal generality of thermodynamic principles across classical and quantum realms,
despite their fundamentally different basis from classical laws of irreversible thermodynamics relying
on local equilibrium assumptions.

\end{abstract}

\maketitle

\section{Introduction}

Thermoelectric devices have garnered significant attention owing to the continual demand for innovative and effective approaches to temperature sensors, heat pumps, and energy conversion~\cite{rowe1995crcbook,francis1999thermoelectric,goldsmid2009introduction,shakouri2009nanoengineered,dubi2011colloquium-heat,mazza2015separation,benenti2017fundamental}. This interest is rooted in the phenomenon of thermoelectricity, where a temperature gradient induces an electric current (Seebeck effect), and a potential gradient induces a heat current (Peltier effect). From a thermodynamic point of view, a non-equilibrium system experiences a distinct set of generalized thermodynamic forces, arising from its simultaneous couplings with different reservoirs~\cite{callen1985book,landi2021irreversible}. The system's response to these external thermodynamic forces is reflected in a corresponding set of generalized thermodynamic fluxes. The concept has been well investigated in classical irreversible thermodynamics, with Onsager's groundbreaking work on the reciprocity principle of thermoelectric phenomena~\cite{onsager1931reciprocal-I,onsager1931reciprocal-II,callen1948the-application}. Traditionally, thermocouples consisting of two different metal wires, are used to observe such reciprocal effects. Only in recent times, experimental research on magnetic metals and insulators, have experienced the emergence of the spin Seebeck effect (SSE), wherein a spin current is generated in response to a thermal gradient~\cite{uchida2008observation,wu2015paramagnetic,zhou2021seebeck}, and conversely, the spin Peltier effect (SPE), involves a spin voltage producing a thermal current~\cite{flipse2014observation,daimon2016thermal,ohnuma2017theory}. The above findings have ignited renewed enthusiasm among researchers to grasp the fundamental aspects of spin caloritronics~\cite{bauer2012spin,boona2014spin,ronetti2016spin,uchida2021transport} and explore practical applications such as waste heat recovery and on-chip refrigeration for future nanoelectronics. As a result, there is a considerable interest in understanding the quantum thermodynamics of nanoscale thermoelectrics through theoretical modelings~\cite{di2008electrical,nazarov2009quantum,ihn2010semiconductor,heikkila2013physics,ren2013predicted,whitney2016thermoelectricity,whitney2018quantum,wang2022cycleflux} and experimental setups involving quantum dot (QD) nanostructures, nanowires, and two-dimensional materials~\cite{houten1992thermoelectric,lee2016thermoelectric,svilans2016experiments,erlingsson2017reversal,patel2020high-thermoelectric,han2020spin,yang2023the-role}.

The quantized energy levels and strong on-site Coulomb interactions among QDs, make them excellent candidates for thermoelectric applications~\cite{esposito2009thermoelectric,nakpathomkun2010thermoelctric,donsa2014double,sothmann2015thermoelectric,whitney2016thermoelectricity,erdman2017thermoelectric,whitney2018quantum,wang2022cycleflux} and various other nanoscale thermal devices~\cite{esposito2012stochastically,thierschmann2015three,jiang2015phonon,zhang2017three,shuvadip2022univarsal}. While the discrete QD spectrum can be fine-tuned via external gate voltages and offers energy-selective transport,  the strong Coulombic interaction between electrons on capacitively coupled QDs can facilitate the transfer of precise amounts of energy from the heat reservoirs. However, the use of QDs as working substances for quantum thermodynamic devices, characterized by a limited number of quantum states, necessitates a completely new understanding of these devices~\cite{whitney2018quantum}. The typical thermalization length being larger than the nanoscale dimension forces these systems to behave in a highly non-trivial manner, and their transport properties cannot be adequately described by the usual Boltzmann transport equation~\cite{datta2005book}, which primarily relies on the local equilibrium assumptions.

On the contrary, the Lindblad master equation, formulated in terms of the density matrix, is used as the preferred tool for examining the thermodynamic properties of the open quantum systems~\cite{breuer2002book,gelbwaser2015thermodynamics,joulain2016quantum,ghosh2017catalysis,potts2019introduction,nikhil2021statistical,gupt2022PRE,shuvadip2022univarsal}. Though it is quite effective in accurately calculating the steady-state currents amid non-equilibrium conditions, it does not reveal any information about the operational principles and the nature of the transport coefficients involved in complex quantum systems. In contrast, network theory in recent years has emerged as a powerful instrument for comprehending non-equilibrium quantum systems~\cite{wang2022cycleflux,gupt2023topranked}. In this framework, dissipative quantum dynamics can be represented as a weighted network featuring nodes and edges~\cite{schnakenberg1976network}. Here, vertices (nodes) signify quantum states, and edges denote non-equilibrium transitions from one quantum state to another, with positive flux rates. Network theory has been applied for many years to explore complex biological phenomena and chemical reactions~\cite{hill1975stochastic,kohler1980thefrequency,ren2017detectable,annwesha2020stochastic}. However, recent work by Wang et al.~\cite{wang2022cycleflux} has drawn huge attention by utilizing network theory to understand the principle working mechanism of quantum thermal devices.  The present authors have extended the technique further to molecular systems to unravel hidden electron transfer pathways in solar cells under strong non-equilibrium conditions~\cite{gupt2023topranked}.

In this paper, we leverage the advantages of network theory to elucidate the operational principles of spin-thermoelectric effects within a three-terminal quantum setup, closely resembling classical thermocouples. We demonstrate how spin and energy currents, obtained from the quantum master equation, are linked to the thermodynamic forces, manifesting spin-Seebeck and spin-Peltier, as thermodynamic cross-effects. Close parallelisms between the microscopic and the macroscopic description of the non-equilibrium system are established via cycle force and cycle fluxes within a basic graph and thermodynamic forces and fluxes of phenomenological laws. The central concept being used here is an expression of the entropy production rate within the framework of the algebraic graph.

The present work is organized as follows: In Sec.~\ref{model-dynamics}, we introduce the basic model of the quantum thermocouple and present the microscopic description using the Lindblad master equation and quantum kinetic Pauli master equation.  We elaborate the basic framework of network theory in the context of spin-thermoelectric effects in Sec.~\ref{network theory} and recover the phenomenological law of irreversible thermodynamics and Onsager's reciprocity in terms of network cycle flux and forces. Operational principles of both spin-Seebeck and spin-Peltier effects are presented in Sec.~\ref{operational principles} and finally, we conclude in Sec.~\ref{conclusion}.

\section{Microscopic Model and Quantum Master equation}\label{model-dynamics}
\begin{figure}[!h]
    \centering   \includegraphics[width=\columnwidth]{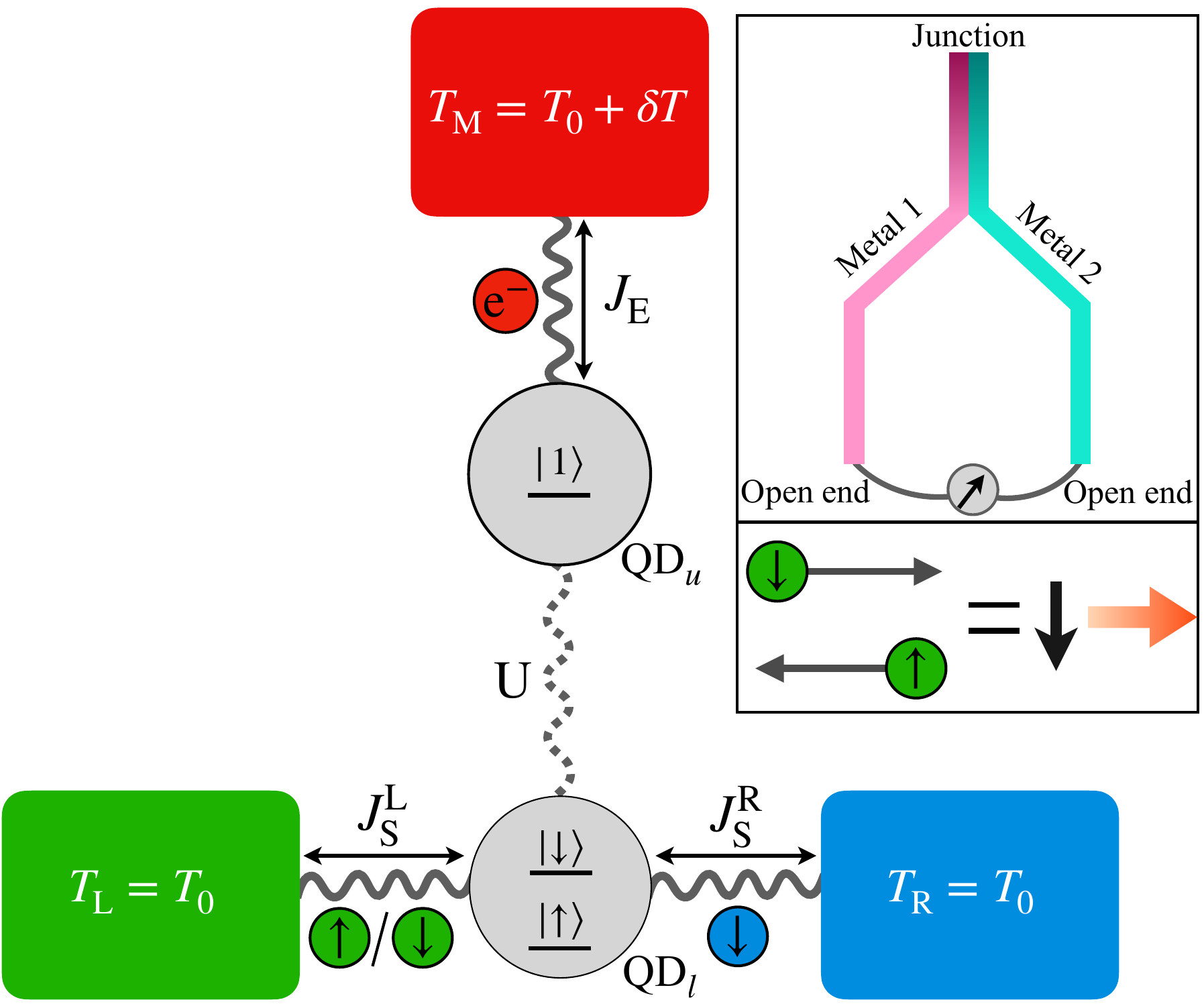}
    \caption{(Inset) Schematic diagram of the classical thermocouple and an illustration of spin current generation. Schematic diagram of a three-terminal Coulomb-coupled QD thermocouple. The lower quantum dot (${\rm QD}_l$) is coupled to the left reservoir i.e. a spinful reservoir (in green) and the right reservoir (in blue) i.e. a magnon bath. Both terminals are kept at equal temperatures and serve as cold ends. The upper quantum dot (${\rm QD}_u$) is coupled to the middle reservoir (in red) i.e. a spinless electron reservoir.  Here, heat is transferred from the middle reservoir which acts as a junction (hot end) and the spin current is across the lower two terminals, analogous to the open ends of a thermocouple.}
    \label{fig1}
\end{figure}

The basic model of a quantum thermocouple consists of two strongly coupled quantum dots (QDs) via Coulomb interaction. The lower quantum dot, denoted as ${\rm QD}_l$, is simultaneously coupled with a spinful free-electron reservoir (on the left) and a magnon bath (on the right), both maintained at an equal temperature ($T_{0}$), as depicted in Fig.~\ref{fig1}. The upper quantum dot, ${\rm QD}_u$, is only coupled with a spinless free-electron reservoir, acting as a junction like in a classical thermocouple [Fig.~\ref{fig1}: top Inset]. The spinful free-electron reservoir comprises spin-polarized electrons with both spin-up ($\uparrow$) and spin-down ($\downarrow$) orientations~\cite{vandaele2017thermal,wang2022cycleflux}. In contrast, the spinless free-electron reservoir in the middle consists of electrons without any distinct spin orientation, and the magnon bath at the right is responsible for inducing spin-flipping of the ${\rm QD}_l$ electrons~\cite{ren2013predicted,wang2022cycleflux,vandaele2017thermal,sothmann2012magnon}. It facilitates the generation of spin current by creating two spin channels propagating in opposite directions, carrying an equal amount of charge current. The spin being an angular momentum, the situation produces a pure spin current without any charge current [Fig.~\ref{fig1}: bottom Inset].

Though two-terminal setups are commonly used to analyze thermoelectricity at the nano-scale, they pose practical challenges in decoupling the heat reservoirs from the electrical circuits~\cite{thierschmann2015three,shakouri2011recent}. By spatially separating the heat reservoir (upper terminal) from the conductor circuit (lower two terminals), the three-terminal model allows us to independently control the direction of spin current using external gate voltages, without affecting the heat current direction. Thus, the three-terminal quantum dot model presented here bears a striking similarity to a classical thermocouple, manifesting both the spin-Seebeck and spin-Peltier effects (SSE and SPE) within a single setting. In SSE, a spin current emerges under the influence of a temperature gradient ($\delta T$), while SPE occurs with the application of a spin bias voltage at the lower terminals, resembling the open ends of a conventional thermocouple. Notably, the difference between the statistical properties of the magnon (bosonic) and electron (fermionic) reservoirs plays a crucial role in generating spin-thermoelectric effects. This distinction can be attributed to the role of dissimilar metal wires in a classical thermocouple, highlighting its significance within the present quantum framework. 

From a practical point of view,  we emphasize that a similar three-terminal setup without any spin consideration has already been experimentally realized by Thierschmann et al.~\cite{thierschmann2015three}, as a nanoscale thermoelectric energy harvester. The device was fabricated from a GaAs/AlGaAs wafer that contains a two-dimensional electron gas (2DEG). Annealed Au/Ge pads were incorporated into the electronic reservoirs to establish reliable electrical connections with the 2DEG. Similar setups can potentially demonstrate spin-thermoelectric effects within a junction system between a magnetic material and a conductor. Generally, a ferrimagnetic insulator like yttrium iron garnet (YIG) and a paramagnetic metal like Pt, are used for this purpose due to their thermally excited dynamics of localized magnetic moments~\cite{uchida2021transport}. Furthermore, the circuit QED magneto-spectroscopic technique can be used to realize spinful states in a double quantum dot~\cite{contamin2023inhomogeneous}, and the spin bias voltage could be measured using the Hanle method~\cite{lebreton2011thermal}, or be converted into an electric voltage through the inverse Hall effect~\cite{uchida2010spin}.

The total Hamiltonian of the entire three-terminal setup is thus given by
\begin{eqnarray}
    H&=&H_{\rm S}+H_{\rm B}+H_{\rm I},\nonumber\\
    H_{\rm S}&=& \sum_{\sigma=\{\uparrow,\downarrow\}}\varepsilon_{l\sigma}{n}_{l\sigma} + \varepsilon_u{n}_u  +\sum_{\sigma=\{\uparrow,\downarrow\}} {\rm U}  {n}_u {n}_{l\sigma},\label{system-H}\\
    H_{\rm B}&=&H_{\rm L}+H_{\rm M}+H_{\rm R},\nonumber\\
            &=&\sum_{\sigma,k}(\epsilon_{{\rm L}\sigma k}-\mu_{{\rm L}\sigma})b^\dagger_{{\rm L}\sigma k}b_{{\rm L}\sigma k}\nonumber\\
            &+&\sum_{k}(\epsilon_{{\rm M} k}-\mu_{\rm M})b^\dagger_{{\rm M} k}b_{{\rm M} k}
            + \sum_{q}\epsilon_{{\rm R} q}a^\dagger_{{\rm R} q}a_{{\rm R} q},\label{bath-H}\\
    H_{\rm I}&=& H_{\rm IL}+H_{\rm IM}+H_{\rm IR},\nonumber\\
            &=& \hbar\sum_{\sigma,k}(t_{{\rm L}k}b^\dagger_{{\rm L}\sigma k} d_{l\sigma}+t^*_{{\rm L}k} d^\dagger_{l\sigma}b_{{\rm L}\sigma k})\nonumber\\
            &+& \hbar\sum_{k}(t_{{\rm M}k}b^\dagger_{{\rm M} k} d_{u}+t^*_{{\rm M}k} d^\dagger_{u}b_{{\rm M} k})\nonumber\\
            &+& \hbar\sum_{q}(g_{{\rm R}q}a^\dagger_{{\rm R} q} d^\dagger_{l\uparrow} d_{l\downarrow}+g^*_{{\rm R}q} d^\dagger_{l\downarrow}d_{l\uparrow}a_{{\rm R} q})\label{System-bath-H}.
\end{eqnarray}
Equation~\eqref{system-H} represents the total system Hamiltonian of the two Coulomb-coupled QDs, where ${\rm U}$ describes the long-range positive Coulomb repulsion energy that permits energy exchange but forbids any particle exchange between the QDs. The operator ${n}_{l\sigma}= d^\dagger_{l\sigma} d_{l\sigma}$ is the number operator for ${\rm QD}_l$, with eigenstates $|\phi_l\rangle=\{|0\rangle, \ket{\uparrow}, \ket{\downarrow}\}$ and corresponding eigenenergies $0$, $\varepsilon_{l\uparrow}$ and $\varepsilon_{l\downarrow}$, respectively, where $d^\dagger_{l\sigma}$ ($d_{l\sigma}$) denotes the electron creation (annihilation) operator with a single particle energy level $\varepsilon_{l\sigma}$, obeying anti-commutation relation $\{d_{l\sigma},d^\dagger_{l\sigma'}\}=\delta_{\sigma\sigma'}$; $\sigma$ being the spin orientation of the electrons. Similarly, $n_u=d^\dagger_u d_u$ is the number operator for ${\rm QD}_u$, with eigenstates $|\phi_u\rangle = {\{|0\rangle, |1\rangle}\}$ and corresponding eigenenergies $0$ and $\varepsilon_{u}$, respectively, where, $d^\dagger_u$ ($d_u$) represents the electron creation (annihilation) operator for ${\rm QD}_u$, with a single particle energy level of $\varepsilon_u$, satisfying the anti-commutation relation $\{d_u,d^\dagger_u\}=1$.

Equation~\eqref{bath-H} describes the total bath Hamiltonian $H_{\rm B}$, wherein $H_{\rm L}$, $H_{\rm M}$ and $H_{\rm R}$ are the respective Hamiltonians for the left (L), middle (M) and right (R) reservoirs. The operators $b^\dagger_{{\rm L}\sigma k}$ ($b^\dagger_{{\rm M} k}$) and $b_{{\rm L}\sigma k}$ ($b_{{\rm M} k}$) represent the creation and annihilation operators of electrons for the L and M baths, where, $\epsilon_{{\rm L}\sigma k}$ and $\mu_{{\rm L}\sigma}$ stand for the energy and chemical potential of electrons corresponding to the spinful fermionic reservoir (L), with $k$ being the continuous wave number (momentum) and $\sigma$ denotes the electron spin. The difference between the chemical potentials $\mu_{\rm L\downarrow}$ and $\mu_{\rm L\uparrow}$ is given by the spin bias voltage i.e., $\Delta\mu_{\rm S}=\mu_{\rm L\downarrow}-\mu_{\rm L\uparrow}$. On the other hand, $\epsilon_{{\rm M} k}$ and $\mu_{\rm M}$ refer to the energy and chemical potential of electrons without any specific spin orientation for the spinless fermionic reservoir (M). For the magnon bath (R), $a^\dagger_{{\rm R}q}$ and $a_{{\rm R}q}$ are the bosonic creation and annihilation operators with the energy $\epsilon_{{\rm R}q}$ and momentum $q$ respectively.

Equation~\eqref{System-bath-H} provides the total system-reservoir interaction Hamiltonian $H_{\rm I}$, where $H_{\rm I\alpha}$ ($\alpha={\rm L,M,R}$) represents the interaction between the system and the $\alpha$-th reservoir. Here the ${\rm QD}_l$ (${\rm QD}_u$) is tunnel-coupled to the L and M reservoir with the tunneling amplitudes $t_{\rm L(M)}$, allowing both particle and energy exchange with the QDs, while ${\rm QD}_l$ is simultaneously coupled to a magnon bath which flips only one spin at a time. Under strong coupling, the eigenstates of $H_{\rm S}$ are determined by the tensor product of the number operator's eigenbasis $|\phi_u \phi_l \rangle$ of the coupled QD system. For convenience, the six microstates of the coupled system ${\{|0\rangle, |1\rangle}\}\otimes \{|0\rangle, \ket{\uparrow}, \ket{\downarrow}\}$, are labeled 
 by $|\mathbb{1}\rangle=|00\rangle$, $|\mathbb{2}\rangle=|10\rangle$, $|\mathbb{3}\rangle=\ket{0\uparrow}$, $|\mathbb{4}\rangle=\ket{0\downarrow}$, $|\mathbb{5}\rangle=\ket{1\uparrow}$, $|\mathbb{6}\rangle=\ket{1\downarrow}$ and their corresponding eigenenergies ($\varepsilon_\mathbb{i}$, $\mathbb{i=1,2,....,6}$) are given by $\varepsilon_\mathbb{1}=0$, $\varepsilon_\mathbb{2}=\varepsilon_u$, $\varepsilon_\mathbb{3}=\varepsilon_{l\uparrow}$, $\varepsilon_\mathbb{4}=\varepsilon_{l\downarrow}$, $\varepsilon_\mathbb{5}=\varepsilon_u +\varepsilon_{l\uparrow} + {\rm U}$ and $\varepsilon_\mathbb{6}=\varepsilon_u +\varepsilon_{l\downarrow} + {\rm U}$ respectively. There are in total nine allowed transitions: The transitions $|\mathbb{1}\rangle \leftrightarrow |\mathbb{3}\rangle$, $|\mathbb{1}\rangle \leftrightarrow |\mathbb{4}\rangle$, $|\mathbb{2}\rangle \leftrightarrow |\mathbb{5}\rangle$ and $|\mathbb{2}\rangle \leftrightarrow |\mathbb{6}\rangle$ are driven by the reservoir L, while the transitions $|\mathbb{1}\rangle \leftrightarrow |\mathbb{2}\rangle$, $|\mathbb{3}\rangle \leftrightarrow |\mathbb{5}\rangle$ and $|\mathbb{4}\rangle \leftrightarrow |\mathbb{6}\rangle$ are induced by the reservoir M, and the transitions $|\mathbb{3}\rangle \leftrightarrow |\mathbb{4}\rangle$ and $|\mathbb{5}\rangle \leftrightarrow |\mathbb{6}\rangle$ are triggered by the bath R.

To calculate the thermal spin ($J_{\rm S}$) and energy current ($J_{\rm E}$) under the SSE and SPE, we first derive the Lindblad quantum master equation of the reduced density matrix $\rho$ for the coupled QDs system under the Born-Markov and Secular (BMS) approximation~\cite{breuer2002book,strasberg2022quantum,gupt2022PRE,gupt2023topranked} (see Appendix \ref{Appendix-A})
\begin{equation}\label{LMEq}
    \frac{d\rho}{dt}=\mathcal{L}_{\rm L}[\rho]+\mathcal{L}_{\rm R}[\rho]+\mathcal{L}_{\rm M}[\rho].
\end{equation}
Here $\mathcal{L}_\alpha$ ($\alpha={\rm L, R, M}$) is the Lindbladian due to the interaction of the quantum system with its $\alpha$-th reservoir. The explicit form of the superoperator $\mathcal{L}$ is given in terms of dissipater
\begin{equation}\label{dissipater}
  \mathcal{D}(C)[\rho]=C \rho C^\dagger-\frac{1}{2}\{\rho,C^\dagger C\},\; C\in \{d_{l\sigma},d_u,d^\dagger_{l\uparrow} d_{l\downarrow}\}, 
\end{equation} 
as follows:
\begin{eqnarray}
    \mathcal{L}_{\rm L}[\rho]&=&\sum_{\sigma=\{\uparrow,\downarrow\}}\mathcal{L}_{{\rm L}\sigma}[\rho],\nonumber\\
    \mathcal{L}_{{\rm L}\sigma}[\rho]&=&\sum_{\{\varepsilon_{{\rm L}\sigma}\}} \gamma_{\rm L}\Big[ f(\varepsilon_{{\rm L}\sigma},\mu_{{\rm L}\sigma},T_{\rm L}) \mathcal{D}(d^\dagger_{l\sigma})[\rho]
    \nonumber\\
   &+& (1-f(\varepsilon_{{\rm L}\sigma},\mu_{{\rm L}\sigma},T_{\rm L}))\mathcal{D}(d_{l\sigma})[\rho] \Big],
   \label{Eqn-Lindbladian-L}
   \\
   \mathcal{L}_{\rm M}[\rho]&=&\sum_{\{\varepsilon_{\rm M}\}} \gamma_{\rm M}\Big[ f(\varepsilon_{\rm M},\mu_{\rm M},T_{\rm M}) \mathcal{D}(d^\dagger_{u})[\rho] \nonumber\\
   &+& (1-f(\varepsilon_{\rm M},\mu_{\rm M},T_{\rm M}))\mathcal{D}(d_{u})[\rho] \Big],
   \label{Eqn-Lindbladian-M}
   \\
   \mathcal{L}_{\rm R}[\rho]&=&\sum_{\{\varepsilon_{\rm R}\}} \gamma_{\rm R}\Big[ n(\varepsilon_{\rm R},T_{\rm R}) \mathcal{D}(d^\dagger_{l\downarrow} d_{l\uparrow})[\rho] \nonumber\\
   &+& (1+n(\varepsilon_{\rm R},T_{\rm R}))\mathcal{D}(d^\dagger_{l\uparrow} d_{l\downarrow})[\rho] \Big].
   \label{Eqn-Lindbladian-R}
\end{eqnarray}
Note that we have implemented the strong coupling formalism to derive the interaction picture master equation presented above~\cite{werlang2014optimal,shuvadip2022univarsal}. Here, the strong coupling refers to the interaction between the two QDs, while the system-reservoir coupling is assumed to be weak, allowing for the safe implementation of the BMS approximation. In Eqs.~\eqref{Eqn-Lindbladian-L}-\eqref{Eqn-Lindbladian-R}, all $\gamma$ values stand for the bare tunneling rates associated with individual processes and depend on the system-reservoir coupling strength through the respective bath spectral function. Lastly, $f(\varepsilon,\mu, T)=[e^{(\varepsilon-\mu)/k_B T}+1]^{-1}$ and $n(\varepsilon, T)=[e^{\varepsilon/k_B T}-1]^{-1}$ are respectively the Fermi-Dirac (FD) and Bose-Einstein (BE) distribution functions with the positive transition energy $\varepsilon$, chemical potential $\mu$ and temperature $T$ associated with the thermal reservoir, where $k_B$ is the Boltzmann constant. Since the Hamiltonian $H_{\rm S}$ in Eq.~\eqref{system-H} is diagonal in the number state eigenbasis of the coupled QDs system, the reduced density matrix $\rho$ of the above Lindblad master equation effectively decouples the diagonal and off-diagonal matrix elements in the eigenbasis of $H_{\rm S}$ \cite{shuvadip2022univarsal}. The diagonal elements of the density matrix $\rho$ signify the occupation probabilities of each microstate and the time evolution is given by 
\begin{eqnarray}
    \frac{dP_\mathbb{1}}{dt}&=&J_\mathbb{12}+J_\mathbb{13}+J_\mathbb{14},\label{P1-dot}\\
    \frac{dP_\mathbb{2}}{dt}&=&J_\mathbb{21}+J_\mathbb{25}+J_\mathbb{26},\\
    \frac{dP_\mathbb{3}}{dt}&=&J_\mathbb{31}+J_\mathbb{34}+J_\mathbb{35},\\
    \frac{dP_\mathbb{4}}{dt}&=&J_\mathbb{41}+J_\mathbb{43}+J_\mathbb{46},\\
    \frac{dP_\mathbb{5}}{dt}&=&J_\mathbb{52}+J_\mathbb{53}+J_\mathbb{56},\\
    \frac{dP_\mathbb{6}}{dt}&=&J_\mathbb{62}+J_\mathbb{64}+J_\mathbb{65}.\label{P6-dot}
\end{eqnarray}
Here $J_\mathbb{ij}$ stands for the net transition rate from state $|\mathbb{j}\rangle$ to $|\mathbb{i}\rangle$ which is given by
\begin{eqnarray} J_\mathbb{ij}&=&k_\mathbb{ij}P_\mathbb{j}-k_\mathbb{ji}P_\mathbb{i},
    \label{transition-rate}\\
    J_\mathbb{ij}&=&-J_\mathbb{ji},\;\quad \mathbb{i,j=1,2,....,6}
\end{eqnarray}
where, $P_\mathbb{i}=\langle i|\rho|i\rangle$ is the population of the $\mathbb{i}$-th eiegenstate and $k_\mathbb{ji}$ ($k_\mathbb{\ket{j}\leftarrow \ket{i}}$) gives the transition probability from microstate $|\mathbb{i}\rangle$ to microstate $|\mathbb{j}\rangle$. The rate expressions $k_\mathbb{ji}$ for all transitions in terms of $\gamma$ and the distribution functions can be summarized as follows:
\begin{eqnarray}\label{transition-rates}
    k_\mathbb{31}&=&\gamma_{\rm L} f(\varepsilon_{l\uparrow},\mu_{{\rm L}\uparrow},T_{\rm L}),\nonumber\\ 
    k_\mathbb{13}&=&\gamma_{\rm L} [1-f(\varepsilon_{l\uparrow},\mu_{{\rm L}\uparrow},T_{\rm L})],\nonumber\\
    k_\mathbb{41}&=&\gamma_{\rm L} f(\varepsilon_{l\downarrow},\mu_{{\rm L}\downarrow},T_{\rm L}),\nonumber\\ 
    k_\mathbb{14}&=&\gamma_{\rm L} [1-f(\varepsilon_{l\downarrow},\mu_{{\rm L}\downarrow},T_{\rm L})],\nonumber\\
    k_\mathbb{52}&=&\gamma_{\rm L} f(\varepsilon_{l\uparrow}+{\rm U},\mu_{{\rm L}\uparrow},T_{\rm L}),\nonumber\\ 
    k_\mathbb{25}&=&\gamma_{\rm L} [1-f(\varepsilon_{l\uparrow}+{\rm U},\mu_{{\rm L}\uparrow},T_{\rm L})],\nonumber\\
    k_\mathbb{62}&=&\gamma_{\rm L} f(\varepsilon_{l\downarrow}+{\rm U},\mu_{{\rm L}\downarrow},T_{\rm L}),\nonumber\\ 
    k_\mathbb{26}&=&\gamma_{\rm L} [1-f(\varepsilon_{l\downarrow}+{\rm U},\mu_{{\rm L}\downarrow},T_{\rm L})],\nonumber\\
    k_\mathbb{21}&=&\gamma_{\rm M} f(\varepsilon_u,\mu_{\rm M},T_{\rm M}),\nonumber\\ 
    k_\mathbb{12}&=&\gamma_{\rm M} [1-f(\varepsilon_u,\mu_{\rm M},T_{\rm M})],\nonumber\\
    k_\mathbb{53}&=&k_\mathbb{64}=\gamma_{\rm M} f(\varepsilon_u+{\rm U},\mu_{\rm M},T_{\rm M}),\nonumber\\ 
    k_\mathbb{35}&=&k_\mathbb{46}=\gamma_{\rm M} [1-f(\varepsilon_u+{\rm U},\mu_{\rm M},T_{\rm M})],\nonumber\\
    k_\mathbb{43}&=&k_\mathbb{65}=\gamma_{\rm R} n(\varepsilon_{l\downarrow}-\varepsilon_{l\uparrow},T_{\rm R}),\nonumber\\
    k_\mathbb{34}&=&k_\mathbb{56}=\gamma_{\rm R} [1+n(\varepsilon_{l\downarrow}-\varepsilon_{l\uparrow},T_{\rm R})].
\end{eqnarray}

Combining Eqs.~\eqref{P1-dot}-\eqref{P6-dot} with Eq.~\eqref{transition-rate}, it is evident that the evolution equations for the microscopic probabilities exhibit linearity with respect to the populations $\{P_\mathbb{i}\}$. As a result, we can cast these equations in the following compact form
\begin{equation}\label{Pauli-MEq}
    \frac{dP_\mathbb{i}}{dt}=\sum^\mathbb{6}_\mathbb{j=1}J_\mathbb{ij}=\sum^\mathbb{6}_\mathbb{j=1} k_\mathbb{ij}P_\mathbb{j}-k_\mathbb{ji}P_\mathbb{i};\quad \mathbb{i\ne j},
\end{equation}
where $\sum^\mathbb{6}_\mathbb{i=1}P_\mathbb{i}=1$. Equation~\eqref{Pauli-MEq} is known as the quantum kinetic Pauli master equation which is ``classical'' in looking but quantum mechanical in content through the transition probabilities $\{k_\mathbb{ij}\}$, determined by the Fermi's golden rule within BMS approximation and the statistical properties of the respective quantum baths~\cite{sinha2011decay,sinha2011quantum}. To obtain the steady-state solution $\Bar{P}_\mathbb{i}$ of Eq.\eqref{Pauli-MEq}, one has to solve the system of linear equations, satisfying the conditions $0\le \Bar{P}_\mathbb{i}\le 1$ and $\sum_\mathbb{i} \Bar{P}_\mathbb{i}=1$. With the help of Eq.~\eqref{Pauli-MEq} and the rate coefficients calculated from the above microscopic picture (Cf.~\eqref{transition-rates}), it is possible to evaluate the steady-state spin and energy currents in terms of the net transition rates (Cf.~Eq.~\eqref{transition-rate}) between the system microstates, where $\{P_\mathbb{i}\}$ get replaced by the steady-state populations $\{\Bar{P}_\mathbb{i}\}$. Following the definition of the spin and energy currents mentioned in Appendix-\ref{Appendix-A}, we obtain the mathematical expression of the steady-state spin current $J_{\rm S}$ which flows from left to right, as
\begin{eqnarray}\label{spin-current}
    J_{\rm S}&=&\frac{1}{2}\Big(\Tr\{d^\dagger_{l\downarrow}d_{l\downarrow}\mathcal{L}_{\rm L\downarrow}[\rho]\}-\Tr\{d^\dagger_{l\uparrow}d_{l\uparrow}\mathcal{L}_{\rm L\uparrow}[\rho]\}\Big),\nonumber\\
    &=& J_\mathbb{34}+ J_\mathbb{56}
    =(k_\mathbb{34}\Bar{P}_\mathbb{4}-k_\mathbb{43}\Bar{P}_\mathbb{3})+(k_\mathbb{56}\Bar{P}_\mathbb{6}-k_\mathbb{65}\Bar{P}_\mathbb{5}),\nonumber\\
    \label{spin-current-Js}
\end{eqnarray}
and the steady-state energy (heat) current $J_{\rm E}$, through the middle reservoir is given by 
\begin{eqnarray}\label{energy-current}
    J_{\rm E}&=&\Tr\{\mathcal{L}_{\rm M}[\rho]H_{\rm S}\}={\rm U}(J_\mathbb{53}+J_\mathbb{64})\nonumber\\
    &=&{\rm U}[(k_\mathbb{53}\Bar{P}_\mathbb{3}-k_\mathbb{35}\Bar{P}_\mathbb{5})+(k_\mathbb{64}\Bar{P}_\mathbb{4}-k_\mathbb{46}\Bar{P}_\mathbb{6})].
\end{eqnarray}
Eq.~\eqref{energy-current} immediately implies that a finite energy current always requires a finite Coulomb interaction energy. However, obtaining the exact analytical solutions for $J_{\rm S}$, $J_{\rm E}$ in terms of steady-state populations [Eqs.~\eqref{spin-current}~and~\eqref{energy-current}] by solving the linear master equation [Eq.~\eqref{Pauli-MEq}] is by no means a trivial task. Secondly, while, one may in principle use exact Eqs.~~\eqref{spin-current-Js} and~\eqref{energy-current} to numerically compute the steady-state spin and energy currents, it does not provide any physical insight into the underlying transport mechanisms leading to SSE and SPE. Nor does it explain how the macroscopic spin and energy currents are related to the thermodynamic forces that give rise to spin-thermoelectric effects as a manifestation of thermodynamic cross-effects.

An alternative yet effective method is to calculate algebraic expressions for steady-state currents through a network or mathematical graph theory~\cite{tutte2001book,balakrishnan2012book}. This also allows us to understand the operational principles of QD-based spin-thermoelectric effects quite easily. In this method, one first constructs a basic graph $\mathbb{G}$ as a \textit{diagrammatic representation} of the right-hand side of Eq.~\eqref{Pauli-MEq}. To extract the principal mechanism from complex transport behaviors, one then decomposes the quantum transition network into cycle trajectories, collects the cycle fluxes using algebraic graph theory, and selects the top-ranked cycle fluxes—i.e., the cycle trajectories with the highest probabilities~\cite{wang2022cycleflux}. In the following section, we illustrate this method in the context of the present problem and establish the connection between the microscopic descriptions of the non-equilibrium system via the basic graph and the macroscopic description of thermolectric phenomena in terms of thermodynamic forces and fluxes, including the celebrated Onsager and Kelvin relations~\cite{callen1985book}. The key concept throughout the entire formalism is the expression of entropy production rate in the framework of graph theory~\cite{landi2021irreversible,schnakenberg1976network}.

\section{Network theory and Reciprocity relation}\label{network theory}

The network or graph theory found its first application in electricity, with Kirchhoff making a pioneering contribution to the understanding of electrical circuits as non-equilibrium systems involving electric current and potential. Since then, graph theory has expanded its horizons and produced a flurry of inspiring early works by  Hill, Kohler, Vollmerhaus, King, and Altman~\cite{hill1975stochastic,kohler1980thefrequency,king1956schematic}, particularly on biophysical and biochemical systems. A vast body of literature is available on this subject~\cite{wu2012efficient,einax2011heterojunction,einax2014network,ren2017detectable,annwesha2020stochastic}; still, Schnakenberg's 1976 review is considered a seminal contribution to this field~\cite{schnakenberg1976network}.

\subsection{Quantum Transition network and Cycle flux analysis}

As an extension of network theory to quantum systems, the notable work of Wang et al.~\cite{wang2022cycleflux} is worth mentioning. They have recently demonstrated that the dissipative quantum dynamics of non-equilibrium transport can be mapped onto networks of quantum state transitions, where nodes or vertices correspond to quantum states, and the connecting lines or edges between two quantum states represent their allowed transitions.
\begin{figure}[!h]
    \centering
    \includegraphics[width=0.9\columnwidth]{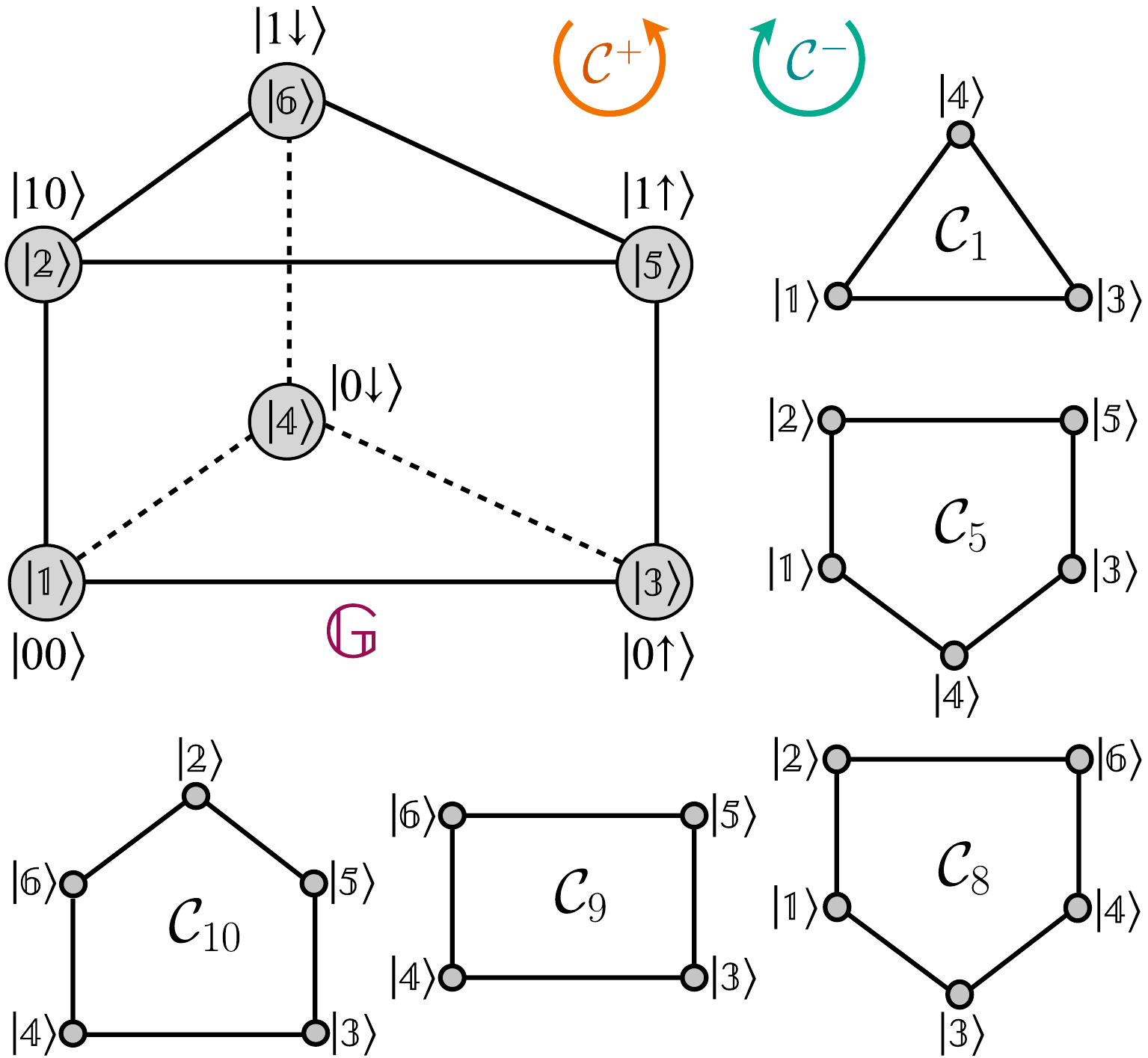}
    \caption{Schematic diagram of the basic graph ($\mathbb{G}$). Subcycles $\{ \mathcal{C}_1,\mathcal{C}_5,\mathcal{C}_8,\mathcal{C}_9,\mathcal{C}_{10}\}$, sharing the common edge ($\ket{\mathbb{3}}\leftrightarrow\ket{\mathbb{4}}$) are used to calculate edge flux $J_\mathbb{43}$ [Cf.~Eq.\eqref{edge-J_43}].}
    \label{fig2}
\end{figure}
In the present case, the diagrammatic representation of the quantum transport processes under the non-equilibrium condition is shown in Fig.~\ref{fig2} in the form of a basic graph ($\mathbb{G}$), where
each node or vertex represents a quantum state $\{|\mathbb{i}\rangle\}$ along with its associated (microscopic) occupation probability $\{P_\mathbb{i}\}$. The transition between adjacent quantum states $|\mathbb{i}\rangle$ and $|\mathbb{j}\rangle$ are depicted by edges. The steady-state population $\Bar{P}_\mathbb{i}$ can then be calculated as
\begin{equation}\label{P-graph-theory}   \Bar{P}_\mathbb{i}=\frac{\Lambda_\mathbb{i}}{\Lambda}, \quad \text{with} \quad 0\le \Bar{P}_\mathbb{i}\le 1 \quad \text{and}\quad  \sum_\mathbb{i} \Bar{P}_\mathbb{i}=1;
\end{equation}
where $\Lambda_\mathbb{i}$ represents the sum of the weight of the spanning trees rooted on the $\mathbb{\ket{i}}$-th state and $\Lambda$ is defined as the sum of the weights of the spanning trees rooted on every individual state $\{\mathbb{\ket{i}}\}$, i.e. $\sum_\mathbb{i}\Lambda_\mathbb{i}$. In the literature, the above method is known as Kirchhoff's theorem~\cite{schnakenberg1976network,kirchhoff1847}. According to this theorem, a spanning tree is a subgraph of $\mathbb{G}$ which includes all the vertices with the minimum number of edges that are always connected but have no {\it circuits} (cyclic sequence of edges or \textit{cycle trajectory}). To construct a spanning tree, one should remove $\nu=e-v+1$ number of edges of the basic graph $\mathbb{G}$, where $e$ and $v$ are the numbers of edges and vertices in $\mathbb{G}$~\cite{schnakenberg1976network}. As a result, all possible spanning trees contain an equal number of vertices and edges.

Under the non-equilibrium condition, each edge represents a transport process and the rate of these transport processes is determined by the net transition rate or edge flux. The steady-state edge flux from a state $\mathbb{\ket{j}}$ and $\mathbb{\ket{i}}$ is defined as 
\begin{equation}\label{steady-state-edge-flux}   J_\mathbb{ij}=k_\mathbb{ij}\Bar{P}_\mathbb{j}-k_\mathbb{ji}\Bar{P}_\mathbb{i},
\end{equation}
where, each edge denotes a pair of transitions with the transition probabilities $k_\mathbb{ij}$ (from $|\mathbb{j}\rangle$ to $|\mathbb{i}\rangle$) and $k_\mathbb{ji}$ (from $|\mathbb{i}\rangle$ to $|\mathbb{j}\rangle$) ~\cite{schnakenberg1976network,wu2012efficient}. 
Measuring edge currents in a graph or network is a challenging task, yet, techniques like scanning tunneling microscopy (STM) can provide insights into edge currents for fabricated nanoscale systems~\cite{chen2021introduction}. Typically, a basic graph $\mathbb{G}$ is comprised of numerous undirected subcycles ($\mathcal{C}$), and each of these subcycles represents a pair of two one-directional circuits [Fig.~\ref{fig2}], namely $\mathcal{C}^+$ (counterclockwise) and $\mathcal{C}^-$ (clockwise)~\cite{kohler1980thefrequency}. Since the circuits are formed by the cyclic sequence of edges within $\mathbb{G}$, the edge flux can be defined in terms of the circuit fluxes~\cite{schnakenberg1976network}, as
\begin{equation}\label{edge-flux}
J_\mathbb{ij}=\sum_\mathcal{C}\mathcal{S}_\mathbb{ij}(\mathcal{C})(J^+_\mathcal{C}-J^-_\mathcal{C})=\sum_\mathcal{C}\mathcal{S}_\mathbb{ij}(\mathcal{C})J_\mathcal{C}.   
\end{equation}
Here $J_\mathcal{C}=J^+_\mathcal{C}-J^-_\mathcal{C}$ denotes the net \textit{cycle flux} wherein $J^+_\mathcal{C}$ and $J^-_\mathcal{C}$ are the circuit fluxes correspond to circuits $\mathcal{C}^+$ and $\mathcal{C}^-$ respectively, with the prefactor $\mathcal{S}_\mathbb{ij}(\mathcal{C})=0,\pm 1$. $\mathcal{S}_\mathbb{ij}(\mathcal{C})=0$ if $\mathcal{C}^+$ and $\mathcal{C}^-$ does not contain the edge $\mathbb{\ket{j}}\rightarrow\mathbb{\ket{i}}$; $\mathcal{S}_\mathbb{ij}(\mathcal{C})=+1$ if the orientation of $\mathcal{C}^+$ ($\mathcal{C}^-$) is along (opposite) to edge $\mathbb{\ket{j}}\rightarrow\mathbb{\ket{i}}$ and $\mathcal{S}_\mathbb{ij}(\mathcal{C})=-1$ if the orientation of $\mathcal{C}^+$ ($\mathcal{C}^-$) is opposite (along) to edge $\mathbb{\ket{j}}\rightarrow\mathbb{\ket{i}}$. For example, the edge flux $J_\mathbb{43}$ ($J_\mathbb{\ket{4}\leftarrow \ket{3}}$) in the basic graph $\mathbb{G}$, can be expressed in terms of the circuit fluxes [Fig.~\ref{fig2}] as 
\begin{equation}\label{edge-J_43} 
J_\mathbb{43}=J^+_{\mathcal{C}_1}-J^-_{\mathcal{C}_1}-J^+_{\mathcal{C}_5}+J^-_{\mathcal{C}_5}+J^+_{\mathcal{C}_8}-J^-_{\mathcal{C}_8}-J^+_{\mathcal{C}_9}+J^-_{\mathcal{C}_9}-J^+_{\mathcal{C}_{10}}+J^-_{\mathcal{C}_{10}}.
\end{equation}
The name ``circuit'' was initially introduced by Kohler and Vollmerhaus~\cite{kohler1980thefrequency} and also termed a ``one-way cycle'' by Hill~\cite{hill1966studies}. However, we prefer to use the term ``circuit'' or ``cycle trajectory'' to avoid confusion with the usual ``cycle''. Hill and Chen provided the physical interpretation for circuit fluxes~\cite{hill1975stochastic}, revealing that these fluxes signify the `frequency' (or rate) of circuit completions along a particular cycle trajectory. To be specific, the circuit flux associated with a one-directional cycle trajectory $\mathcal{C^{\pm}}$ is given by
\begin{equation}\label{circuit-flux-graph-theory}
J^\pm_{\mathcal{C}}=\Pi^\pm_{\mathcal{C}}\frac{\Lambda_\mathcal{C}}{\Lambda}.
\end{equation}
Here, $\Pi^\pm_{\mathcal{C}}$ denotes the weight factor which is determined by the product of the transition rates along the circuit $\mathcal{C^{\pm}}$. For example, the clockwise cycle trajectory $\mathcal{C}^-_1(\mathbb{\ket{1}}\rightarrow\mathbb{\ket{4}}\rightarrow\mathbb{\ket{3}}\rightarrow\mathbb{\ket{1}})$ [Fig.~\ref{fig2}] has the weight factor $\Pi^-_{\mathcal{C}_1}=k_\mathbb{13}k_\mathbb{34}k_\mathbb{41}$, where, $\Lambda_\mathcal{C}$ represents the sum of the weight of the spanning trees rooted on cycle $\mathcal{C}$ and $\Lambda=\sum_{\mathbb{i}}\Lambda_{\mathbb{i}}$. 
Now, there are a total of 22 paired cycle trajectories, or 11 subcycles, for our basic graph $\mathbb{G}$, which are as follows:
\begin{eqnarray}
    \mathcal{C}_1 &:& |\mathbb{1}\rangle\leftrightarrow|\mathbb{3}\rangle\leftrightarrow|\mathbb{4}\rangle\leftrightarrow|\mathbb{1}\rangle \nonumber\\
    \mathcal{C}_2 &:& |\mathbb{2}\rangle\leftrightarrow|\mathbb{5}\rangle\leftrightarrow|\mathbb{6}\rangle\leftrightarrow|\mathbb{2}\rangle \nonumber\\
    \mathcal{C}_3 &:& |\mathbb{1}\rangle\leftrightarrow|\mathbb{3}\rangle\leftrightarrow|\mathbb{5}\rangle\leftrightarrow|\mathbb{2}\rangle\leftrightarrow|\mathbb{1}\rangle \nonumber\\
    \mathcal{C}_4 &:& |\mathbb{1}\rangle\leftrightarrow|\mathbb{3}\rangle\leftrightarrow|\mathbb{5}\rangle\leftrightarrow|\mathbb{6}\rangle\leftrightarrow|\mathbb{2}\rangle\leftrightarrow|\mathbb{1}\rangle \nonumber\\
    \mathcal{C}_5 &:& |\mathbb{1}\rangle\leftrightarrow|\mathbb{4}\rangle\leftrightarrow|\mathbb{3}\rangle\leftrightarrow|\mathbb{5}\rangle\leftrightarrow|\mathbb{2}\rangle\leftrightarrow|\mathbb{1}\rangle\nonumber\\
    \mathcal{C}_6 &:& |\mathbb{1}\rangle\leftrightarrow|\mathbb{4}\rangle\leftrightarrow|\mathbb{6}\rangle\leftrightarrow|\mathbb{2}\rangle\leftrightarrow|\mathbb{1}\rangle \nonumber\\
    \mathcal{C}_7 &:& |\mathbb{1}\rangle\leftrightarrow|\mathbb{4}\rangle\leftrightarrow|\mathbb{6}\rangle\leftrightarrow|\mathbb{5}\rangle\leftrightarrow|\mathbb{2}\rangle\leftrightarrow|\mathbb{1}\rangle \nonumber\\
    \mathcal{C}_8 &:& |\mathbb{1}\rangle\leftrightarrow|\mathbb{3}\rangle\leftrightarrow|\mathbb{4}\rangle\leftrightarrow|\mathbb{6}\rangle\leftrightarrow|\mathbb{2}\rangle\leftrightarrow|\mathbb{1}\rangle \nonumber\\
    \mathcal{C}_9 &:& |\mathbb{3}\rangle\leftrightarrow|\mathbb{5}\rangle\leftrightarrow|\mathbb{6}\rangle\leftrightarrow|\mathbb{4}\rangle\leftrightarrow|\mathbb{3}\rangle \nonumber\\
    \mathcal{C}_{10} &:& |\mathbb{2}\rangle\leftrightarrow|\mathbb{6}\rangle\leftrightarrow|\mathbb{4}\rangle\leftrightarrow|\mathbb{3}\rangle\leftrightarrow|\mathbb{5}\rangle\leftrightarrow|\mathbb{2}\rangle \nonumber\\
    \mathcal{C}_{11} &:& |\mathbb{1}\rangle\leftrightarrow|\mathbb{3}\rangle\leftrightarrow|\mathbb{5}\rangle\leftrightarrow|\mathbb{6}\rangle\leftrightarrow|\mathbb{4}\rangle\leftrightarrow|\mathbb{1}\rangle.
\end{eqnarray}
Hence, enumerating a large number of spanning trees rooted at each individual state, as well as for cycles, poses a formidable challenge. This difficulty becomes more pronounced with the increasing size of the basic graph.
To bypass this problem, we utilize the generalized matrix-tree theorem from algebraic graph~\cite{wang2022cycleflux,gupt2023topranked} by rewriting the master equation in the following form~\cite{keizer1972onthesolutions}: $\dot{\rm \textbf{P}}=-{\rm \textbf{MP}}$, where ${\rm \textbf{P}}=\{P_\mathbb{1}, P_\mathbb{2}, P_\mathbb{3}, P_\mathbb{4}, P_\mathbb{5}, P_\mathbb{6}\}$ is a column matrix and ${\rm \textbf{M}}$ is a square matrix, given by
\begin{widetext}
\begin{eqnarray}\label{laplacianMatrix}
{\rm \textbf{M}}=\left[ {\begin{array}{cccccc} k_{\mathbb{21}}+k_{\mathbb{31}}+k_{\mathbb{41}} & -k_{\mathbb{12}} & -k_{\mathbb{13}} & -k_{\mathbb{14}} & 0 & 0 \\ 
   -k_{\mathbb{21}} & k_{\mathbb{12}}+k_{\mathbb{52}}+k_{\mathbb{62}} & 0 & 0 & -k_{\mathbb{25}} & -k_{\mathbb{26}} \\
  -k_{\mathbb{31}} & 0 & k_{\mathbb{13}}+k_{\mathbb{43}}+k_{\mathbb{53}} & -k_{\mathbb{34}} & -k_{\mathbb{35}} & 0 \\ 
   -k_{\mathbb{41}} & 0 & -k_{\mathbb{43}} & k_{\mathbb{14}}+k_{\mathbb{34}}+k_{\mathbb{64}} & 0 & -k_{\mathbb{46}} \\
   0 & -k_{\mathbb{52}} & -k_{\mathbb{53}} & 0 & k_{\mathbb{25}}+k_{\mathbb{35}}+k_{\mathbb{65}} & -k_{\mathbb{56}} \\
   0 & -k_{\mathbb{62}} & 0 & -k_{\mathbb{64}} & -k_{\mathbb{65}} & k_{\mathbb{26}}+k_{\mathbb{46}}+k_{\mathbb{56}} \\
  \end{array} } \right].
\end{eqnarray}
\end{widetext}
Equation~\eqref{laplacianMatrix} is known as the Laplacian or transition matrix of the weighted graph $\mathbb{G}$. Furthermore, in accordance with the matrix tree theorem, it is possible to compute both the numerator and denominator of Eqs.~\eqref{P-graph-theory} and \eqref{circuit-flux-graph-theory} as the determinants of the reduced transition matrix. 
\begin{figure}[!h]
	\centering   \includegraphics[width=\columnwidth]{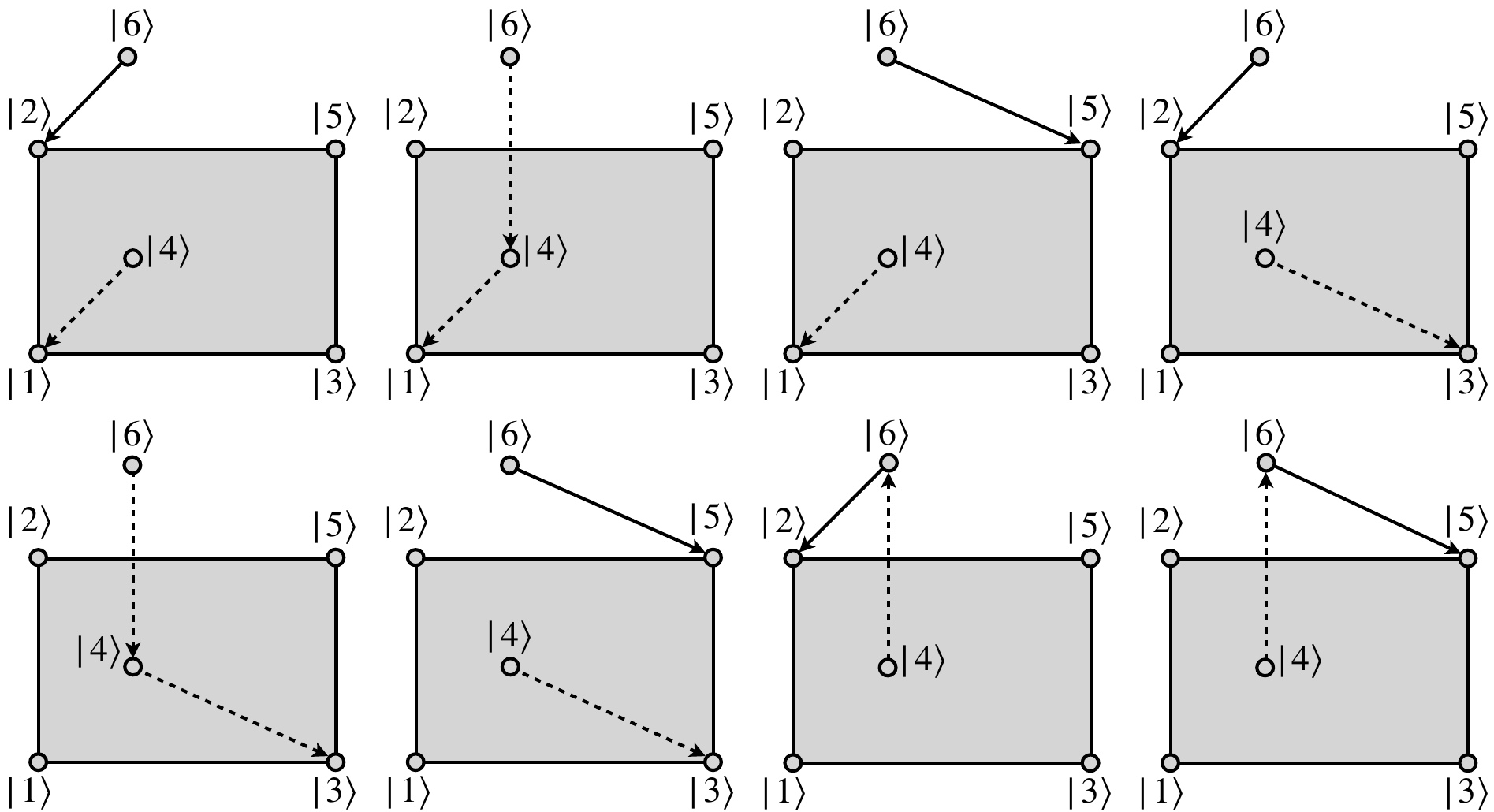}
	\caption{Spanning trees  rooted on cycle $\mathcal{C}_3$ (shaded region) of the basic graph.}
	\label{fig3}
\end{figure}
For instance, $\Lambda_\mathbb{i}$ is related to ${\rm \textbf{M}}[\mathbb{i}, \mathbb{i}]$ which can be obtained by removing the $\mathbb{i}$-th row and column of the Laplacian matrix ${\rm \textbf{M}}$. Similarly, $\Lambda_\mathcal{C}$ is identical to the $\det({\rm \textbf{M}}[\mathcal{C}, \mathcal{C}])$, obtained by deleting rows and columns belonging to cycle $\mathcal{C}$ of the transition matrix $\textbf{M}$. This directly leads to a simple algebraic expression of the steady-state population 
\begin{equation}\label{Populations-MT} \Bar{P}_\mathbb{i}=\frac{\det({\rm \textbf{M}}[\mathbb{i}; \mathbb{i}])}{\sum_\mathbb{i}\det({\rm \textbf{M}}[\mathbb{i}; \mathbb{i}])},
\end{equation}
and the one-directional circuit flux associated with circuits $\mathcal{C}^{\pm}$ in the following form
\begin{equation}\label{circuit-flux-MT}
J^\pm_\mathcal{C}=\Pi^\pm_\mathcal{C}\frac{\det({\rm \textbf{M}}[\mathcal{C}; \mathcal{C}])}{\sum_\mathbb{i}\det({\rm \textbf{M}}[\mathbb{i}; \mathbb{i}])},
\end{equation}
where ${\sum_\mathbb{i}\det({\rm \textbf{M}}[\mathbb{i}; \mathbb{i}])}=\Lambda$. As an example, the sum of the weights of spanning trees rooted on cycle $\mathcal{C}_3$ in terms of the reduced determinant of the original Laplacian matrix ${\rm \textbf{M}}$ is obtained by deleting rows and columns $\mathbb{i (1,2,3,5)} \in \mathcal{C}_3$ [Fig.~\ref{fig3}]. So, the principle minor ${\rm \textbf{M}}[\mathcal{C}_3, \mathcal{C}_3]$ or ${\rm \textbf{M}}[1,3,5,2; 1,3,5,2]$ and its determinant takes the form of
\begin{eqnarray}
    {\rm \textbf{M}}[\mathcal{C}_3, \mathcal{C}_3]=\left[ {\begin{array}{ccc}  k_\mathbb{14}+k_\mathbb{34}+k_\mathbb{64}  & -k_\mathbb{46} \\
       -k_\mathbb{64} & k_\mathbb{26}+k_\mathbb{46}+k_\mathbb{56} 
    \end{array}} \right],
\end{eqnarray}
and
\begin{eqnarray}\label{Lambda_C3}
    \det({\rm \textbf{M}}[\mathcal{C}_3, \mathcal{C}_3])=k_\mathbb{14}k_\mathbb{26}+k_\mathbb{14}k_\mathbb{46}+k_\mathbb{14}k_\mathbb{56}+k_\mathbb{34}k_\mathbb{26} \nonumber\\    +k_\mathbb{34}k_\mathbb{46}+k_\mathbb{34}k_\mathbb{56}+k_\mathbb{26}k_\mathbb{64}+k_\mathbb{56}k_\mathbb{64}.
\end{eqnarray}
respectively. Equation~\eqref{Lambda_C3} contains a sum of eight terms, indicating that there are eight possible spanning trees rooted on cycle $\mathcal{C}_3$. Each term represents the weight of a spanning tree in which all the weighted edges are directed towards the cycle $\mathcal{C}_3$ [Fig.~\ref{fig3}]. Finally, from Eq.~\eqref{circuit-flux-MT}, it is clear that we can calculate the circuit fluxes $J^\pm_{\mathcal{C}}$ for each stochastic cycle trajectory with the orientation either clockwise or counterclockwise and efficiently rank out the top-ranked circuit fluxes. Next, we will show how the microscopic details of cycle and circuit fluxes help us understand the SSE and SPE, connecting spin and energy currents to macroscopic thermodynamic forces in the phenomenological laws of irreversible thermodynamics.

\subsection{Onsager Relation}\label{Onsager-relation}

The first step on our way from a microscopic to a macroscopic description is to establish an expression for entropy production rate, the key quantity in understanding any irreversible processes~\cite{schnakenberg1976network,landi2021irreversible}. To start with, we consider the von Neumann entropy 
\begin{equation}
    \mathcal{S}=-k_B\sum_\mathbb{i} P_\mathbb{i}\ln P_\mathbb{i},
\end{equation}
in the framework of our discrete-state quantum transition network, characterized by its microscopic probability $\{P_\mathbb{i}\}$. Thus, the time evolution of $\mathcal{S}$ is given by
\begin{equation}\label{dSbydt}
    \frac{d\mathcal{S}}{dt}=-k_B\sum_\mathbb{i} \frac{dP_\mathbb{i}}{dt}\ln P_\mathbb{i}.
\end{equation}
With the help of the quantum kinetic Pauli master equation ~\eqref{Pauli-MEq}, one can rewrite Eq.~\eqref{dSbydt} as
\begin{eqnarray}
    \frac{d\mathcal{S}}{dt}
    &=& \frac{1}{2}k_B\sum_\mathbb{i,j} J_\mathbb{ij}\ln\Big(\frac{P_\mathbb{j}}{P_\mathbb{i}}\Big).
    \label{dS/dt}
\end{eqnarray}
Now, following Schnakenberg's suggestion~\cite{schnakenberg1976network}, we split Eq.~\eqref{dS/dt} into two parts, 
\begin{equation}\label{entropy-dot}
    \frac{d\mathcal{S}}{dt}=\dot{\Phi}(t) + \dot{\sigma}(t),
\end{equation}
where we identify the first term $\dot{\Phi}(t)$ as the entropy flux rate
\begin{equation}\label{entropy-flux}
    \dot{\Phi}(t)=-\frac{1}{2}k_B\sum_\mathbb{i,j} J_\mathbb{ij}\ln\Big(\frac{k_\mathbb{ij}}{k_\mathbb{ji}}\Big),
\end{equation}
which arises from the interaction between the system and its surroundings. The second term $\dot{\sigma}(t)$ is the total entropy production rate
\begin{eqnarray}\label{sigma-with-edge-flux}
    \dot{\sigma}(t)
    &=&\frac{1}{2}k_B\sum_\mathbb{i,j} (k_\mathbb{ij}P_\mathbb{j}-k_\mathbb{ji}P_\mathbb{i})\ln\Big(\frac{k_\mathbb{ij}P_\mathbb{j}}{k_\mathbb{ji}P_\mathbb{i}}\Big).
\end{eqnarray}
Equations~\eqref{sigma-with-edge-flux} may appear a little artificial at first glance, and a natural question to be raised at this point is whether Eq.~\eqref{sigma-with-edge-flux} has anything to do with the entropy production of the phenomenological irreversible thermodynamics, which needs be expressed as a bilinear form of the macroscopic thermodynamic forces and fluxes. It must be emphasized that neither $\dot{\mathcal{S}}$ in Eq.~\eqref{entropy-dot} nor $\dot{\Phi}$ in Eq.~\eqref{entropy-flux} are necessarily positive, but only $\dot{\sigma}(t) \geq 0$, since it takes a form $(a-b)\ln(a/b) \geq 0$ [Cf.~Eq.~\eqref{sigma-with-edge-flux}]. This is indeed true since the total entropy production rate ($\dot{\sigma}$) of any system must be always positive~\cite{schnakenberg1976network,landi2021irreversible}. Thus, it turns out that Eq.~\eqref{sigma-with-edge-flux} satisfies the basic criteria for the entropy production rate. Under the steady-state condition, there is no change in the entropy of the system which implies~\cite{schnakenberg1976network,landi2021irreversible}
\begin{equation}\label{st-sigma}
    \dot{\sigma}=-\dot{\Phi}(t)=\frac{1}{2}k_B\sum_\mathbb{i,j} (k_\mathbb{ij}\Bar{P}_\mathbb{j}-k_\mathbb{ji}\Bar{P}_\mathbb{i})\ln\Big(\frac{k_\mathbb{ij}}{k_\mathbb{ji}}\Big).
\end{equation}
Using Eqs.~\eqref{steady-state-edge-flux}~and~\eqref{edge-flux}, one may rewrite the above equation in terms of the circuit and cycle fluxes as~\cite{schnakenberg1976network}, 
\begin{eqnarray}\label{st-sigma-in-Jc}
    \dot{\sigma}&=&k_B\sum_\mathcal{C} (J^+_\mathcal{C}\mathcal{A}^+_{\mathcal{C}}+J^-_\mathcal{C}\mathcal{A}^-_{\mathcal{C}}) \nonumber\\
    &=& k_B\sum_\mathcal{C} (J^+_\mathcal{C} - J^-_\mathcal{C})\mathcal{A}^+_{\mathcal{C}}=k_B\sum_\mathcal{C} J_\mathcal{C}\mathcal{X}_{\mathcal{C}},
\end{eqnarray}
where $\mathcal{X}_\mathcal{C}=\mathcal{A}^+_{\mathcal{C}}=\ln(\Pi^+_\mathcal{C}/\Pi^-_\mathcal{C})=-\mathcal{A}^-_{\mathcal{C}}$ is called the the cycle affinity. For a given cycle $\mathcal{C}$, it measures the imbalance or asymmetry between the transition rates along two opposite cycle trajectories $\mathcal{C}^{\pm}$ and hence qualifies as a thermodynamic force~\cite{ohga2023thermodynamic}. This is because, when $\mathcal{X}_\mathcal{C}=0$, it implies $J_\mathcal{C}=0$, resulting in equal circuit fluxes in both directions, i.e., $J^+_\mathcal{C}=J^-_\mathcal{C}$. Equation~\eqref{st-sigma-in-Jc}, expressed in terms of cycle fluxes and cycle forces, can thus be regarded as a microscopic or stochastic version of the phenomenological Onsager relation~\cite{landi2021irreversible}.

Moreover, we find from Eq.~\eqref{circuit-flux-MT} that the ratio of $J^\pm_\mathcal{C}$ is equal to the ratio of weight factors $\Pi^\pm_\mathcal{C}$ for each cycle, which, in turn, is determined by the ratio of the product of the transitions rates along circuits $\mathcal{C}^{\pm}$ and can be computed in terms of externally controllable, macroscopic physical quantities $T_{0}$, $\delta T$ and $\Delta\mu_{\rm S}$, as follows (see Appendix~\ref{Appendix-B}): 
\begin{eqnarray}
    \frac{J^+_{\mathcal{C}_1}}{J^-_{\mathcal{C}_1}}&=&\frac{\Pi^+_{\mathcal{C}_1}}{\Pi^-_{\mathcal{C}_1}}=
    e^{-{\Delta\mu_{\rm S}}/{k_B T_0}},
    \label{JC1-ratio}\\
    \frac{J^+_{\mathcal{C}_2}}{J^-_{\mathcal{C}_2}}&=&\frac{\Pi^+_{\mathcal{C}_2}}{\Pi^-_{\mathcal{C}_2}}=
    e^{-{\Delta\mu_{\rm S}}/{k_B T_0}} ,\\
    \frac{J^+_{\mathcal{C}_3}}{J^-_{\mathcal{C}_3}}&=&\frac{\Pi^+_{\mathcal{C}_3}}{\Pi^-_{\mathcal{C}_3}}=
    e^{{{\rm U}\delta T}/{k_B T_0(T_0 + \delta T)}},\\
    \frac{J^+_{\mathcal{C}_4}}{J^-_{\mathcal{C}_4}}&=&\frac{\Pi^+_{\mathcal{C}_4}}{\Pi^-_{\mathcal{C}_4}}=
    e^{{{\rm U}\delta T}/{k_B T_0(T_0 + \delta T)}}e^{-{\Delta\mu_{\rm S}}/{k_B T_0}},\\
    \frac{J^+_{\mathcal{C}_5}}{J^-_{\mathcal{C}_5}}&=&\frac{\Pi^+_{\mathcal{C}_5}}{\Pi^-_{\mathcal{C}_5}}=
    e^{{{\rm U}\delta T}/{k_B T_0(T_0 + \delta T)}}e^{{\Delta\mu_{\rm S}}/{k_B T_0}},\\
    \frac{J^+_{\mathcal{C}_6}}{J^-_{\mathcal{C}_6}}&=&\frac{\Pi^+_{\mathcal{C}_6}}{\Pi^-_{\mathcal{C}_6}}=
    e^{{{\rm U}\delta T}/{k_B T_0(T_0 + \delta T)}},\\
    \frac{J^+_{\mathcal{C}_7}}{J^-_{\mathcal{C}_7}}&=&\frac{\Pi^+_{\mathcal{C}_7}}{\Pi^-_{\mathcal{C}_7}}=
    e^{{{\rm U}\delta T}/{k_B T_0(T_0 + \delta T)}}e^{{\Delta\mu_{\rm S}}/{k_B T_0}},\\
    \frac{J^+_{\mathcal{C}_8}}{J^-_{\mathcal{C}_8}}&=&\frac{\Pi^+_{\mathcal{C}_8}}{\Pi^-_{\mathcal{C}_8}}=
    e^{{{\rm U}\delta T}/{k_B T_0(T_0 + \delta T)}}e^{-{\Delta\mu_{\rm S}}/{k_B T_0}},\\   
    \frac{J^+_{\mathcal{C}_9}}{J^-_{\mathcal{C}_9}}&=&\frac{\Pi^+_{\mathcal{C}_9}}{\Pi^-_{\mathcal{C}_9}}=1,
    \label{JC9-ratio}\\
    \frac{J^+_{\mathcal{C}_{10}}}{J^-_{\mathcal{C}_{10}}}&=&\frac{\Pi^+_{\mathcal{C}_{10}}}{\Pi^-_{\mathcal{C}_{10}}}=
    e^{{\Delta\mu_{\rm S}}/{k_B T_0}}, \\
    \frac{J^+_{\mathcal{C}_{11}}}{J^-_{\mathcal{C}_{11}}}&=&\frac{\Pi^+_{\mathcal{C}_{11}}}{\Pi^-_{\mathcal{C}_{11}}}=
    e^{-{\Delta\mu_{\rm S}}/{k_B T_0}},
    \label{JC11-ratio}
\end{eqnarray}
In order to derive Eqs.~\eqref{JC1-ratio}-\eqref{JC11-ratio}, one makes use of Eq.~\eqref{transition-rates}, where the explicit form of the distribution functions are governed by the quantum statistical properties of the respective thermal reservoirs (see Appendix~\ref{Appendix-B} for details). The advantages in writing the above set of equations as a ratio of the circuits fluxes rely on the fact that if the external biases $\Delta \mu_{\rm S}$ and $\delta T$ are zero on the r.h.s of Eqs.~\eqref{JC1-ratio}-\eqref{JC11-ratio}, then regardless of the magnitudes the circuit fluxes, corresponding cycle can't contribute the spin and energy currents. Therefore, one can infer the cycle fluxes associated with the subcycles $\{\mathcal{C}_1, \mathcal{C}_2, \mathcal{C}_{10}, \mathcal{C}_{11} \}$ are controlled by the spin bias voltage $\Delta\mu_{\rm S}$ and hence can only contribute to the spin current $J_{\rm S}$. Whereas the net cycle fluxes associated with the subcycles $\mathcal{C}_3$ and $\mathcal{C}_6$ are dependent only on the temperature gradient $\delta T$ and thereby contributing solely to the energy current $J_{\rm E}$. Note that ${\rm U}$ term is only associated with $\delta T$, and not $\Delta \mu_s$. This observation implies that a non-zero Coulomb interaction ${\rm U}$ is always necessary for a finite energy current, as indicated in conjunction with Eq.~\eqref{energy-current}. However, there are few subcycles $\{\mathcal{C}_4, \mathcal{C}_5, \mathcal{C}_7, \mathcal{C}_8\}$ and their conjugate fluxes are governed by both $\delta T$ as well as $\Delta\mu_{\rm S}$ and therefore can contribute to both $J_{\rm S}$ and $J_{\rm E}$. Indeed, these are the four cycles that are responsible for the spin-thermoelectric cross-effects of SSE and SPE as we will demonstrate in Sec.~\ref{operational principles}. Note that the net cycle flux associated with $\mathcal{C}_9$ is identically zero because the circuit fluxes in both directions (clockwise and counterclockwise) are equal as evident from Eq.\eqref{JC9-ratio}. As a result, we can use Eq.~\eqref{edge-flux} to rewrite macroscopic spin and energy current expressions [Eqs. \eqref{spin-current} and \eqref{energy-current}] in terms of the microscopic circuit and cycle fluxes as follows:
\begin{eqnarray}
    J_{\rm S}&=&-(J^+_{\mathcal{C}_1}-J^-_{\mathcal{C}_1})-(J^+_{\mathcal{C}_2}-J^-_{\mathcal{C}_2})-(J^+_{\mathcal{C}_4}-J^-_{\mathcal{C}_4})\nonumber\\
    &+&(J^+_{\mathcal{C}_5}-J^-_{\mathcal{C}_5})+(J^+_{\mathcal{C}_7}-J^-_{\mathcal{C}_7})-(J^+_{\mathcal{C}_8}-J^-_{\mathcal{C}_8})\nonumber\\
    &+&(J^+_{\mathcal{C}_{10}}-J^-_{\mathcal{C}_{10}})-(J^+_{\mathcal{C}_{11}}-J^-_{\mathcal{C}_{11}}) 
    \label{SC-in-terms-of-circuit-flux}\\
    &=&-J_{\mathcal{C}_1}-J_{\mathcal{C}_2}-J_{\mathcal{C}_4}+J_{\mathcal{C}_5}+J_{\mathcal{C}_7}-J_{\mathcal{C}_8}\nonumber\\
    &+&J_{\mathcal{C}_{10}}-J_{\mathcal{C}_{11}},
    \label{SC-in-terms-of-cycle-flux}
\end{eqnarray}
\begin{eqnarray}
    J_{\rm E}&=&{\rm U}[(J^+_{\mathcal{C}_3}-J^-_{\mathcal{C}_3})+(J^+_{\mathcal{C}_4}-J^-_{\mathcal{C}_4})+(J^+_{\mathcal{C}_5}-J^-_{\mathcal{C}_5})\nonumber\\
    &+&(J^+_{\mathcal{C}_6}-J^-_{\mathcal{C}_6})+(J^+_{\mathcal{C}_7}-J^-_{\mathcal{C}_7})+(J^+_{\mathcal{C}_8}-J^-_{\mathcal{C}_8})]
    \label{JC-in-terms-of-circuit-flux}\\
    &=& {\rm U}[J_{\mathcal{C}_3}+J_{\mathcal{C}_4}+J_{\mathcal{C}_5}+J_{\mathcal{C}_6}+J_{\mathcal{C}_7}+J_{\mathcal{C}_8}].
    \label{EC-in-terms-of-cycle-flux}
\end{eqnarray}
Similar to Eqs.~\eqref{spin-current}~and~\eqref{energy-current}, the above set of equations are the most general ones, however, the latter has an advantage over the previous set of equations. Eqs.~\eqref{JC-in-terms-of-circuit-flux}~and~\eqref{EC-in-terms-of-cycle-flux} can be expressed in terms of macroscopic forces, facilitating the connection between SSE and SPE as a manifestation of thermodynamic cross-effects. In order to identify the phenomenological forces, we 
substitute Eqs.~\eqref{JC1-ratio}-\eqref{JC11-ratio} into Eq.~\eqref{st-sigma-in-Jc}, to write the entropy production rate as a sum over the associated thermodynamic forces and fluxes as (Appendix~\ref{Appendix-B})
\begin{eqnarray}\label{sigma}
    \dot{\sigma}&=&{\rm U}[J_{\mathcal{C}_3}+J_{\mathcal{C}_4}+J_{\mathcal{C}_5}+J_{\mathcal{C}_6}+J_{\mathcal{C}_7}+J_{\mathcal{C}_8}]\frac{\delta T}{T_0(T_0+\delta T)} \nonumber\\
    &-& [J_{\mathcal{C}_1}+J_{\mathcal{C}_2}+J_{\mathcal{C}_4}-J_{\mathcal{C}_5}-J_{\mathcal{C}_7}+J_{\mathcal{C}_8}
    -J_{\mathcal{C}_{10}}+J_{\mathcal{C}_{11}}]\frac{\Delta\mu_{\rm S}}{T_0}\nonumber\\
    &=& J_{\rm E}\Big[\frac{1}{T_0}-\frac{1}{(T_0+\delta T)}\Big] + J_{\rm S}\Big[\frac{\mu_{\rm L\downarrow}}{T_0}-\frac{\mu_{\rm L\uparrow}}{T_0}\Big]\nonumber\\
    &=& J_{\rm E} \mathcal{X}_{\rm E} + J_{\rm S} \mathcal{X}_{\rm S},
\end{eqnarray}
where $\mathcal{X}_{\rm E}$ and $\mathcal{X}_{\rm S}$ are identified as conjugate forces corresponding to the energy current $J_{\rm E}$, and spin current $J_{\rm S}$, respectively. If we compare Eq.~\eqref{st-sigma-in-Jc}~and ~\eqref{sigma}, we observe that in both cases, the entropy production rate $\dot{\sigma}$ is the product of the fluxes and forces: In Eq.\eqref{st-sigma-in-Jc}, $\dot{\sigma}$ is in terms of the microscopic fluxes ($J_\mathcal{C}$) and its conjugate forces ($\mathcal{X}_\mathcal{C}$), i.e., cycle affinities whereas in Eq.\eqref{sigma}, $\dot{\sigma}$ is in terms of the macroscopic fluxes (like the flow of spin and energy current) and the associated phenomenological forces ($\mathcal{X}_{\rm E}$ and $\mathcal{X}_{\rm S}$). This is one of our central results, showcasing the recovery of the phenomenological thermodynamic law of entropy production in terms of generalized thermodynamic forces and fluxes derived from the microscopic dynamical framework of the master equation employing network cycle flux and forces. For small external bias $\delta T$ and $\Delta\mu_{\rm S}$, we can simplify Eq.~\eqref{sigma} in the following form
\begin{eqnarray}\label{sigma-approx}
    \dot{\sigma}\approx J_{\rm E} \left(\frac{\delta T}{{T_0}^2}\right)  + J_{\rm S} \left(\frac{\Delta\mu_{\rm S}}{T_0}\right),
\end{eqnarray}
which is in accordance with the linear dependence of the entropy production rate on the relevant thermodynamic forces. Similarly, the spin and the energy currents within the linear response regime can be approximated as follows:
\begin{widetext}
\begin{eqnarray}
    J_{\rm S}&\approx& {\rm U}(-J^-_{\mathcal{C}_4}+J^+_{\mathcal{C}_5}+J^-_{\mathcal{C}_7}-J^+_{\mathcal{C}_8})\Bigg(\frac{\delta T}{k_B{T_0}^2}\Bigg) + (J^-_{\mathcal{C}_1}+J^-_{\mathcal{C}_2}+J^-_{\mathcal{C}_4}+J^+_{\mathcal{C}_5}+J^-_{\mathcal{C}_7}+J^+_{\mathcal{C}_8}+J^-_{\mathcal{C}_{10}}+J^-_{\mathcal{C}_{11}})\Bigg(\frac{\Delta\mu_{\rm S}}{k_B T_0}\Bigg),
    \label{approx-JS}\\
    J_{\rm E}&\approx& {\rm U^2}(J^-_{\mathcal{C}_3}+J^-_{\mathcal{C}_4}+J^+_{\mathcal{C}_5}+J^-_{\mathcal{C}_6}+J^-_{\mathcal{C}_7}+J^+_{\mathcal{C}_8})\Bigg(\frac{\delta T}{k_B{T_0}^2}\Bigg) + {\rm U}(-J^-_{\mathcal{C}_4}+J^+_{\mathcal{C}_5}+J^-_{\mathcal{C}_7}-J^+_{\mathcal{C}_8})\Bigg(\frac{\Delta\mu_{\rm S}}{k_B T_0}\Bigg).
    \label{approx-JE}
\end{eqnarray}
\end{widetext} 
Comparing Eqs.~\eqref{approx-JS}~and~\eqref{approx-JE} with the phenomenological linear law of irreversible thermodynamics
\begin{eqnarray}
    J^{\rm ph}_{\rm S}= L_{\rm SE}\Bigg(\frac{\delta T}{k_B{T_0}^2}\Bigg) + L_{\rm SS}\Bigg(\frac{\Delta\mu_{\rm S}}{k_B T_0}\Bigg),\label{J_S}\\
    J^{\rm ph}_{\rm E}= L_{\rm EE}\Bigg(\frac{\delta T}{k_B{T_0}^2}\Bigg) + L_{\rm ES}\Bigg(\frac{\Delta\mu_{\rm S}}{k_B T_0}\Bigg),
\end{eqnarray}
we identify the Onsager transport coefficients ($L$'s) in terms of the microscopic circuit fluxes obtained from the network theory  
\begin{eqnarray}
    L_{\rm SE}&=& {\rm U}(-J^-_{\mathcal{C}_4}+J^+_{\mathcal{C}_5}+J^-_{\mathcal{C}_7}-J^+_{\mathcal{C}_8}) \equiv L_{\rm ES},\label{reciprocal-relation}\\
    L_{\rm SS}&=& (J^-_{\mathcal{C}_1}+J^-_{\mathcal{C}_2}+J^-_{\mathcal{C}_4}+J^+_{\mathcal{C}_5}+J^-_{\mathcal{C}_7} \nonumber\\
    &+&J^+_{\mathcal{C}_8}+J^-_{\mathcal{C}_{10}}+J^-_{\mathcal{C}_{11}}),\label{eqn_L_SS}\\
    L_{\rm EE}&=& {\rm U^2}(J^-_{\mathcal{C}_3}+J^-_{\mathcal{C}_4}+J^+_{\mathcal{C}_5}+J^-_{\mathcal{C}_6}+J^-_{\mathcal{C}_7}+J^+_{\mathcal{C}_8}).\label{eqn_L_EE}
\end{eqnarray}
Equation~\eqref{reciprocal-relation} encapsulates the essence of the Onsager reciprocity relation. Here we derive this relation by applying the quantum kinetic Pauli master equation within the framework of network theory. It reveals that the BMS quantum master equation is not a mere description of the dissipative dynamics of the open quantum system; rather, it reproduces the reciprocity relation of the linear law of irreversible thermodynamics, which obeys due to the time-reversal symmetry of the stationary fluctuations. Here instead, it follows from the properties of the network circuit fluxes between the forward (counterclockwise) and reverse (clockwise) cycle trajectories.

Now, our aim is to establish the relationship between the coefficients of the spin-Seebeck and the spin-Peltier effects. Under the zero spin current condition, i.e., $J_{\rm S}=0$, we obtain from Eq.~\eqref{J_S}
\begin{eqnarray}
&&L_{\rm SE}\Bigg(\frac{\delta T}{k_B{T_0}^2}\Bigg) + L_{\rm SS}\Bigg(\frac{\Delta\mu_{\rm S}}{k_B T_0}\Bigg)=0, \nonumber\\
    {\rm or}, \quad \kappa&&\equiv\Bigg(\frac{\Delta\mu_{\rm S}}{\delta T}\Bigg)_{J_{\rm S}=0}=-\frac{1}{T_0}\Bigg(\frac{L_{\rm SE}}{L_{\rm SS}}\Bigg) \label{Seebeck-Coefficient} \\
    \kappa&&=-\frac{\rm U}{T_0}\Big(-J^-_{\mathcal{C}_4}+J^+_{\mathcal{C}_5}+J^-_{\mathcal{C}_7}-J^+_{\mathcal{C}_8}\Big)/\Big(J^-_{\mathcal{C}_1}+J^-_{\mathcal{C}_2} \nonumber\\
    &&+J^-_{\mathcal{C}_4}+J^+_{\mathcal{C}_5}+J^-_{\mathcal{C}_7} +J^+_{\mathcal{C}_8}+J^-_{\mathcal{C}_{10}}+J^-_{\mathcal{C}_{11}}\Big).
    \label{Seebeck-Coefficient-1}
\end{eqnarray}
Here $\kappa=({\Delta\mu_{\rm S}}/{\delta T})_{J_{\rm S}=0}$ is the spin-Seebeck coefficient or spin-thermoelectric power, defined as the change in the spin bias voltage per unit change of temperature. Similarly, we may define the spin-Peltier coefficient as
\begin{eqnarray}
    \vartheta &&= -\Bigg(\frac{J_{\rm E}}{J_{\rm S}} \Bigg)_{{\delta T}=0}=-\frac{L_{\rm ES}}{L_{\rm SS}},\label{Peltier-coefficient} \\
    &&= -{\rm U}\Big(-J^-_{\mathcal{C}_4}+J^+_{\mathcal{C}_5}+J^-_{\mathcal{C}_7}-J^+_{\mathcal{C}_8}\Big)/\Big(J^-_{\mathcal{C}_1}+J^-_{\mathcal{C}_2} \nonumber\\
    &&+J^-_{\mathcal{C}_4}+J^+_{\mathcal{C}_5}+J^-_{\mathcal{C}_7} +J^+_{\mathcal{C}_8}+J^-_{\mathcal{C}_{10}}+J^-_{\mathcal{C}_{11}}\Big).
    \label{Peltier-coefficient-1}
\end{eqnarray}
On the face of it, both Eqs.~\eqref{Seebeck-Coefficient}~and~\eqref{Peltier-coefficient}, appear exactly the same as their classical counterparts, although their basis is completely different in classical and quantum cases. Using Eqs. \eqref{Seebeck-Coefficient-1} and \eqref{Peltier-coefficient-1}, we immediately conclude that the classic \textit{Kelvin relation} 
\begin{eqnarray}
    T_0\Bigg(\frac{\Delta\mu_{\rm S}}{\delta T}\Bigg)_{J_{\rm S}=0}&&=-\Bigg(\frac{J_{\rm E}}{J_{\rm S}} \Bigg)_{{\delta T}=0}=-\frac{L_{\rm ES}}{L_{\rm SS}}=-\frac{L_{\rm SE}}{L_{\rm SS}}, \nonumber
\end{eqnarray}
\begin{equation}\label{kelvin-relation}
{\rm or},\quad  T_0 \kappa= \vartheta.
\end{equation}
equally holds for quantum thermocouples, connecting the two thermoelectric effects, namely SSE and SPE. This is the hallmark of thermodynamics with its universal generality. The generality prevails in the sense that all the thermodynamic relations retain their forms in both classical and quantum settings, with the only variation being in specific expressions that are used to articulate them.

The superiority of graph theory lies in its ability to provide a clear and comprehensive understanding of the fundamental mechanisms underlying spin-thermoelectric effects, which can otherwise be a daunting task to infer from transition states alone. For instance, it is immediately clear from Eqs.~\eqref{JC1-ratio}-\eqref{JC11-ratio}, which cycles are only contributing in $J_{\rm E}$ or $J_{\rm S}$ and which cycles are responsible for the thermodynamic cross-effects. The connection between the stochastic description of non-equilibrium systems in terms of cycle fluxes and cycle forces, and the macroscopic depiction of thermoelectric phenomena characterized by thermodynamic forces and fluxes, demands further attention. This synthesis results in the retrieval of the classic Onsager and Kelvin relations in the most intriguing way. Such integration facilitates the transition from a microscopic perspective to a macroscopic phenomenon, thereby enhancing our understanding and allowing us to explore the intricate interplay between various factors contributing to spin-thermoelectric effects, as discussed below.

\section{Operational Principles}\label{operational principles}
To this end, we delve into the operational principles underlying the spin-Seebeck and spin-Peltier effects.

\subsection{Spin-Seebeck effect}
\begin{figure}[!h]
    \centering \includegraphics[width=\columnwidth]{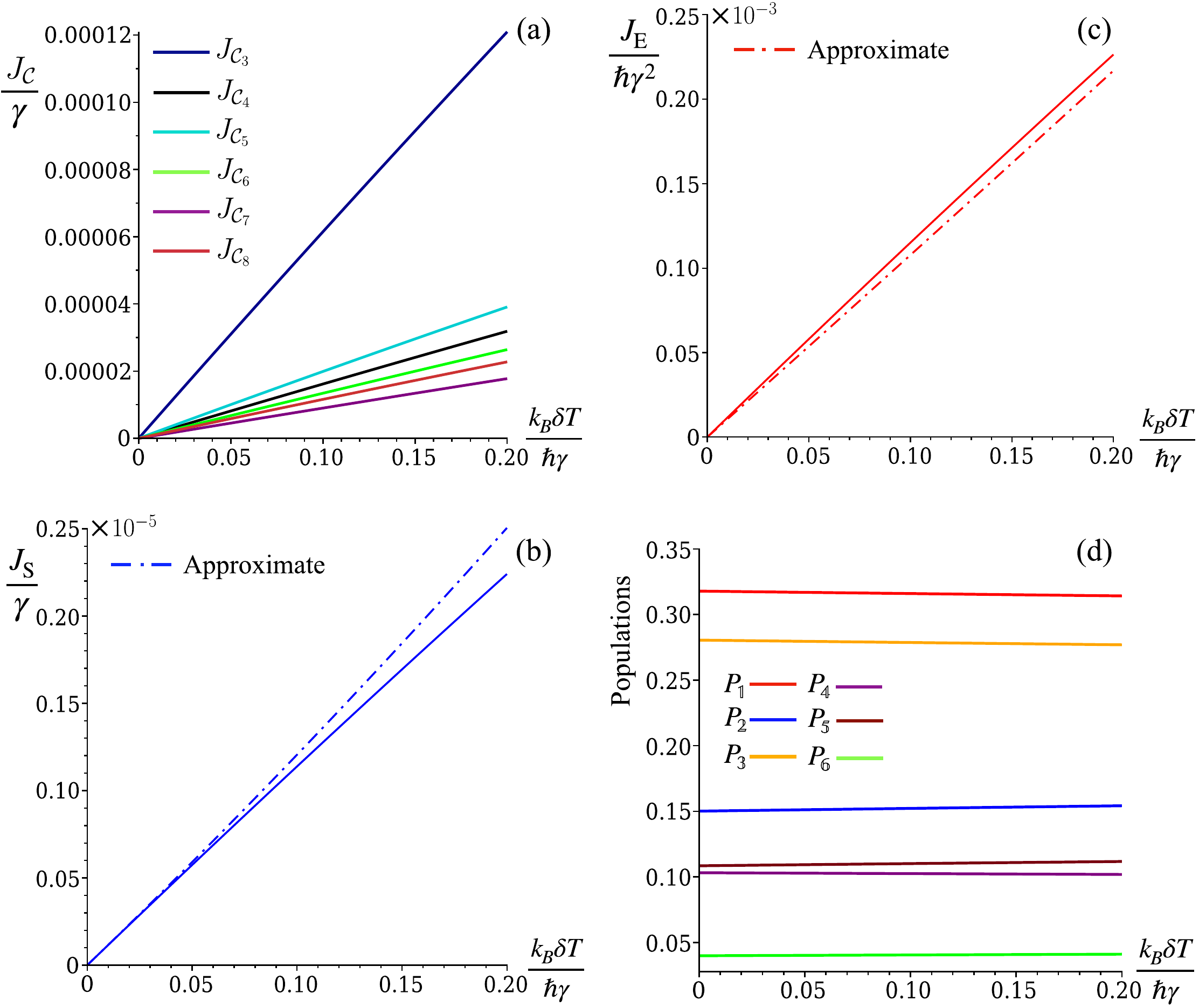}
    \caption{Spin-Seebeck effect: All contributing (a) cycle fluxes (b) populations (c) spin and (d) energy currents are plotted against dimensionless thermal energy $k_B\delta T/\hbar\gamma$. The parameters used are as follows: $\gamma_{\rm L}=\gamma_{\rm M}=\gamma_{\rm R}=\gamma$, ${\rm U}=0.8\hbar\gamma$, $k_BT_0= 4\hbar\gamma$, $\varepsilon_{ l\downarrow}=4.5\hbar\gamma$, $\varepsilon_{l\uparrow}=0.5\hbar\gamma$, $\varepsilon_{u}=3\hbar\gamma$, and $\mu_{\rm L\downarrow}=\mu_{\rm L\uparrow}=0$, $\mu_{\rm M}=0$.}
    \label{SSE-plots}
\end{figure}

We observe the SSE when there is no spin bias voltage $\Delta\mu_{\rm S}=0$, and a spin current is generated due to a temperature difference $\delta T$ between the upper and the lower terminals of the device. For $\Delta\mu_{\rm S}=0$, Eq.\eqref{SC-in-terms-of-circuit-flux} reduces to
\begin{eqnarray}
    J_{\rm S}&=&-(J^+_{\mathcal{C}_4}-J^-_{\mathcal{C}_4}) + (J^+_{\mathcal{C}_5}-J^-_{\mathcal{C}_5})+(J^+_{\mathcal{C}_7}-J^-_{\mathcal{C}_7})\nonumber\\
    &-&(J^+_{\mathcal{C}_8}-J^-_{\mathcal{C}_8})\nonumber\\
    &=&-J_{\mathcal{C}_4}+J_{\mathcal{C}_5}+J_{\mathcal{C}_7}-J_{\mathcal{C}_8},
    \label{JS-for-SSE}
\end{eqnarray}
while the $J_{\rm E}$ is still governed by Eq.~\eqref{EC-in-terms-of-cycle-flux}. As a result, we identify that cycle fluxes corresponding to subcycles $\mathcal{C}_3$, $\mathcal{C}_4$, $\mathcal{C}_5$, $\mathcal{C}_6$, $\mathcal{C}_7$, and $\mathcal{C}_8$ contribute to $J_{\rm E}$, while $\mathcal{C}_4$, $\mathcal{C}_5$, $\mathcal{C}_7$, and $\mathcal{C}_8$ facilitate $J_{\rm S}$. All the contributing cycle fluxes, energy, and spin currents, in dimensionless units, along with all six microstate populations, are plotted in Fig.~\ref{SSE-plots} w.r.t the dimensionless temperature gradient $\delta T$. Although all six cycles assist $J_{\rm E}$, the primary contribution comes from $\mathcal{C}_3$, classified as the highest-rank cycle with a nonzero contribution. In contrast, all the cycles appearing in $J_{\rm S}$ have equal magnitudes but are lower in rank compared to $\mathcal{C}_3$ [Fig.~\ref{SSE-plots}a]. Consequently, the dimensionless spin current $J_{\rm S}$ observed in the SSE, which is a linear combination of the four contributing cycles, is two orders of magnitude less than the dimensionless energy current $J_{\rm E}$ [Figs.~\ref{SSE-plots}b and \ref{SSE-plots}c]. This is a typical feature of SSE, where a weak spin current is generated due to the temperature difference between the hot and cold terminals of the device. The numerical results plotted in Figs.~\ref{SSE-plots}b~and~\ref{SSE-plots}c, in a way, verify this assertion. Upon setting $\Delta\mu_{\rm S}=0$ in Eqs. \eqref{approx-JS} and \eqref{approx-JE}, approximate expressions for $J_{\rm S}$ and $J_{\rm E}$ in the linear response regime are reduced to
\begin{equation}
   J_{\rm S}\approx L_{\rm SE}\Bigg(\frac{\delta T}{k_B{T_0}^2}\Bigg),\label{J-S-apprx-SSE}
\end{equation}
and
\begin{equation}
    J_{\rm E}\approx L_{\rm EE}\Bigg(\frac{\delta T}{k_B{T_0}^2}\Bigg),\label{J-E-apprx-SSE}
\end{equation}
respectively, where $L_{\rm SE}$ and $L_{\rm EE}$ are given by Eqs. \eqref{reciprocal-relation} and \eqref{eqn_L_EE} respectively. We observe that Eqs.~\eqref{J-S-apprx-SSE}~and~\eqref{J-E-apprx-SSE} closely follow the general Eqs.~\eqref{JS-for-SSE}~and~\eqref{EC-in-terms-of-cycle-flux} [solid lines in Figs.~\ref{SSE-plots}b~and~\ref{SSE-plots}c], but they start to deviate [dash-dot lines] for large values of the temperature gradient. This is because Eqs.~\eqref{SC-in-terms-of-circuit-flux}-\eqref{sigma} are exact expressions in terms of circuit and cycle fluxes under linear response regime; whereas, Eqs.~\eqref{sigma-approx}-\eqref{approx-JE} are approximated ones, satisfying phenomenological law of irreversible thermodynamics in its standard form, where we are approximating the energy force $\mathcal{X}_{\rm E}=\delta T/T_{0}(T_{0}+\delta T)$, by replacing with $\delta T/T^2_{0}$. Hence, there is close agreement between the two results for smaller $\delta T$, but the difference increases for larger $\delta T$.  

Finally, we note that cycles ($\mathcal{C}_4$, $\mathcal{C}_5$, $\mathcal{C}_7$, and $\mathcal{C}_8$) involving spin-flip processes contribute to $J_{\rm S}$. For example, consider the dynamic steps of $\mathcal{C}^+_4$: starting from the most populated state $\ket{00}$ as shown in Fig.\ref{SSE-plots}(d), the system sequentially transitions to $\ket{0\uparrow}$ (where one spin-up electron tunnels from the left reservoir into the lower QD), then to $\ket{1\uparrow}$ (where one electron tunnels from the middle reservoir into the upper QD). The third step involves a spin-flip process $\ket{1\uparrow}\rightarrow\ket{1\downarrow}$ by absorbing one magnon supplied by the right reservoir. Subsequently, one spin-up electron tunnels into the left reservoir ($\ket{1\downarrow}\rightarrow\ket{10}$), and finally, the system returns to its initial state $\ket{00}$ by releasing one electron to the middle reservoir. Thus, at the end of the full cycle, an integer spin-1 is transferred from the left spinful electron reservoir to the right magnon bath. Similarly, the clockwise circuit $\mathcal{C}^-_4$ represents the reverse process, and both cycle trajectories additively contribute to the spin current expression [Eq.\eqref{JS-for-SSE}] in the SSE. The same holds true for other contributing cycle trajectories mentioned in Eq.~\eqref{JS-for-SSE}.

\subsection{Spin-Peltier effect}
We observe the SPE in a scenario when $\delta T=0$ and an energy current is generated due to a non-zero spin bias voltage. Putting $\delta T=0$ in Eq~\eqref{EC-in-terms-of-cycle-flux}, we obtain
\begin{eqnarray}
       J_{\rm E}&=&{\rm U}[(J^+_{\mathcal{C}_4}-J^-_{\mathcal{C}_4})+(J^+_{\mathcal{C}_5}-J^-_{\mathcal{C}_5})+(J^+_{\mathcal{C}_7}-J^-_{\mathcal{C}_7})\nonumber\\ &+& (J^+_{\mathcal{C}_8}-J^-_{\mathcal{C}_8})] \\
    &=& {\rm U}[J_{\mathcal{C}_4}+J_{\mathcal{C}_5}+J_{\mathcal{C}_7}+J_{\mathcal{C}_8}],
    \label{JE-for-SPE}
\end{eqnarray}
and the same Eq.\eqref{SC-in-terms-of-circuit-flux} can be used to calculate $J_{\rm S}$. 
\begin{figure}[!h]
    \centering
    \includegraphics[width=\columnwidth]{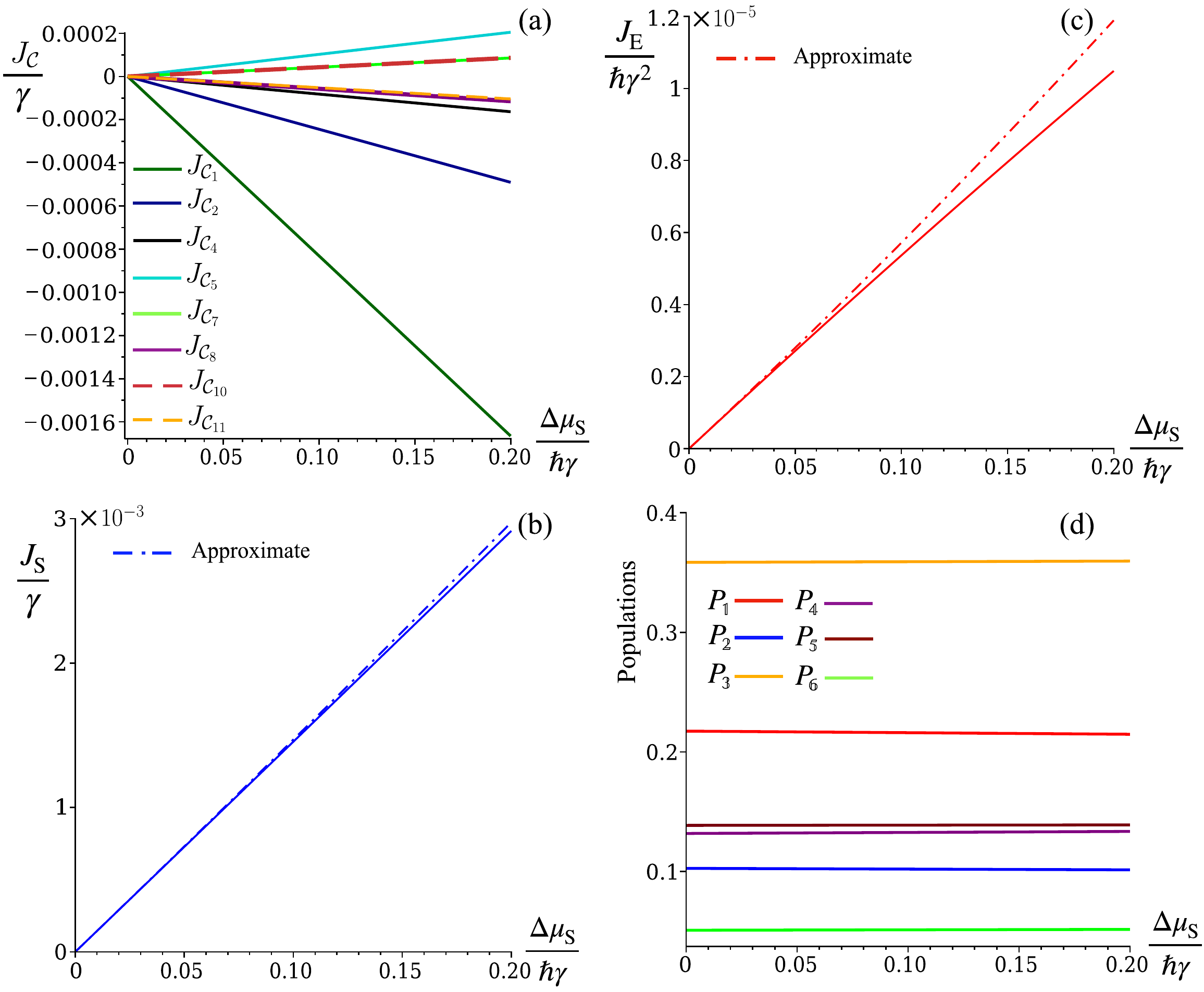}
    \caption{Spin-Peltier effect: All contributing (a) cycle fluxes (b) populations (c) spin and (d) energy currents are plotted against dimensionless thermal energy $\Delta\mu_{\rm S}/\hbar\gamma$. The parameters used are as follows: $\gamma_{\rm L}=\gamma_{\rm M}=\gamma_{\rm R}=\gamma$, ${\rm U}=0.8\hbar\gamma$, $k_BT_0= 4\hbar\gamma$, $\varepsilon_{ l\downarrow}=4.5\hbar\gamma$, $\varepsilon_{l\uparrow}=0.5\hbar\gamma$, $\varepsilon_{u}=3\hbar\gamma$, and $\mu_{\rm L\uparrow}=2.5\hbar\gamma$, $\mu_{\rm L\downarrow}=\mu_{\rm L\uparrow}+\Delta\mu_{\rm S}$, $\delta T=0$,$\mu_{\rm M}=0$.}
    \label{SPE-J-vs-dmu-plot}
\end{figure}
As a result, cycle fluxes corresponding to cycles $\mathcal{C}_1$, $\mathcal{C}_2$, $\mathcal{C}_4$, $\mathcal{C}_5$, $\mathcal{C}_7$, $\mathcal{C}_8$, $\mathcal{C}_{10}$, and $\mathcal{C}_{11}$ attribute to the spin current, while only four cycles $\mathcal{C}_4$, $\mathcal{C}_5$, $\mathcal{C}_7$, and $\mathcal{C}_8$ contribute to the energy current. All supporting cycle fluxes, spin and energy currents in dimensionless units, and the population of each eigenstate are plotted in Fig.\ref{SPE-J-vs-dmu-plot} against the dimensionless spin bias voltage $\Delta\mu_{\rm S}$. In this case, the major contribution to $J_{\rm S}$ is coming from cycle $\mathcal{C}_1$ which is classified as the top-ranked cycle with a nonzero contribution. On the other hand, all the cycles contributing to $J_{\rm E}$ are lower ranked cycles relative to $\mathcal{C}_1$ with comparable magnitudes [Fig.\ref{SPE-J-vs-dmu-plot}a]. As a result, the dimensionless energy current $J_{\rm E}$ in SPE, is two orders of magnitude smaller than the dimensionless spin current $J_{\rm S}$ [Fig.\ref{SPE-J-vs-dmu-plot}b and \ref{SPE-J-vs-dmu-plot}c]. This is a characteristic of SPE, where a weak thermal current is generated due to the spin bias voltage applied across the two ends of the cold terminals. The situation is exactly the reverse of that of SSE and the numerical results plotted in Figs.~\ref{SPE-J-vs-dmu-plot}b and \ref{SPE-J-vs-dmu-plot}c, agree with this observation. This simply indicates that the current conjugate to its thermodynamic forces is always greater than the currents corresponding to the other thermodynamic forces. Upon substituting $\delta T=0$ in Eqs.\eqref{approx-JS} and \eqref{approx-JE}, approximate expressions for $J_{\rm E}$ and $J_{\rm S}$ in the linear response regime take the form of
\begin{equation}
    J_{\rm E}\approx L_{\rm ES}\Bigg(\frac{\Delta\mu_{\rm S}}{k_B T_0}\Bigg),\label{J-E-apprx}
\end{equation}
and
\begin{equation}
    J_{\rm S}\approx L_{\rm SS}\Bigg(\frac{\Delta\mu_{\rm S}}{k_B T_0}\Bigg),\label{J-S-apprx}
\end{equation}
where $L_{\rm ES}$ and $L_{\rm SS}$ are given by Eqs. \eqref{reciprocal-relation} and \eqref{eqn_L_SS}, respectively. Note that Eq.~\eqref{J-S-apprx} closely follow the solid line [Cf.~\eqref{SC-in-terms-of-circuit-flux}] in Fig.~\ref{SPE-J-vs-dmu-plot}b, while the energy current $J_{\rm E}$ deviates gradually from the solid line for higher $\Delta \mu_{\rm S}$ [Fig.~\ref{SPE-J-vs-dmu-plot}c]. This is because, Eqs.~\eqref{approx-JS}-\eqref{approx-JE} obey the Onsager reciprocity, fulfilling $L_{\rm {SE}}=L_{\rm ES}$. As a consequence, the $J_{\rm E}$ expression in Eq.~\eqref{J-E-apprx} is influenced more than the spin current expression [Eq.~\eqref{J-S-apprx}], upon approximating energy force $\mathcal{X}_{\rm E}$ by $\delta T /T^{2}_{0}$ in Eqs.~\eqref{approx-JS}-\eqref{approx-JE}.

Finally, we note that in both Figs.~\ref{SSE-plots}d~and~\ref{SPE-J-vs-dmu-plot}d, the change in the populations exhibits little variation w.r.t $\delta T$ and $\Delta \mu_{s}$, suggesting a linear response of operation close to equilibrium. Had it been operated far from equilibrium, as recently explored in molecular photocells~\cite{gupt2023topranked} driven by solar radiations, it would have resulted in population inversions and other interesting effects. As mentioned earlier, each cycle $\mathcal{C}$ represents two paired circuits, i.e., $\mathcal{C}^+$ and $\mathcal{C}^-$. In Fig.\ref{SPE-J-vs-dmu-plot}a, $J_{\mathcal{C}}<0$  implies that the flux corresponding to the counterclockwise circuit $\mathcal{C}^+$ ($J^+_{\mathcal{C}}$) is less than the flux corresponding to clockwise circuit $\mathcal{C}^-$ ($J^-_{\mathcal{C}}$). Finally, we emphasize that it is the same set of four cycles ($\mathcal{C}_4$, $\mathcal{C}_5$, $\mathcal{C}_7$, and $\mathcal{C}_8$) that not only produces a weak spin current in SSE but also accounts for generating a weak energy current in SPE.

\subsection{SSE and SPE: As thermodynamic cross-effect}

\begin{figure}
    \centering
    \includegraphics[width=0.9\columnwidth]{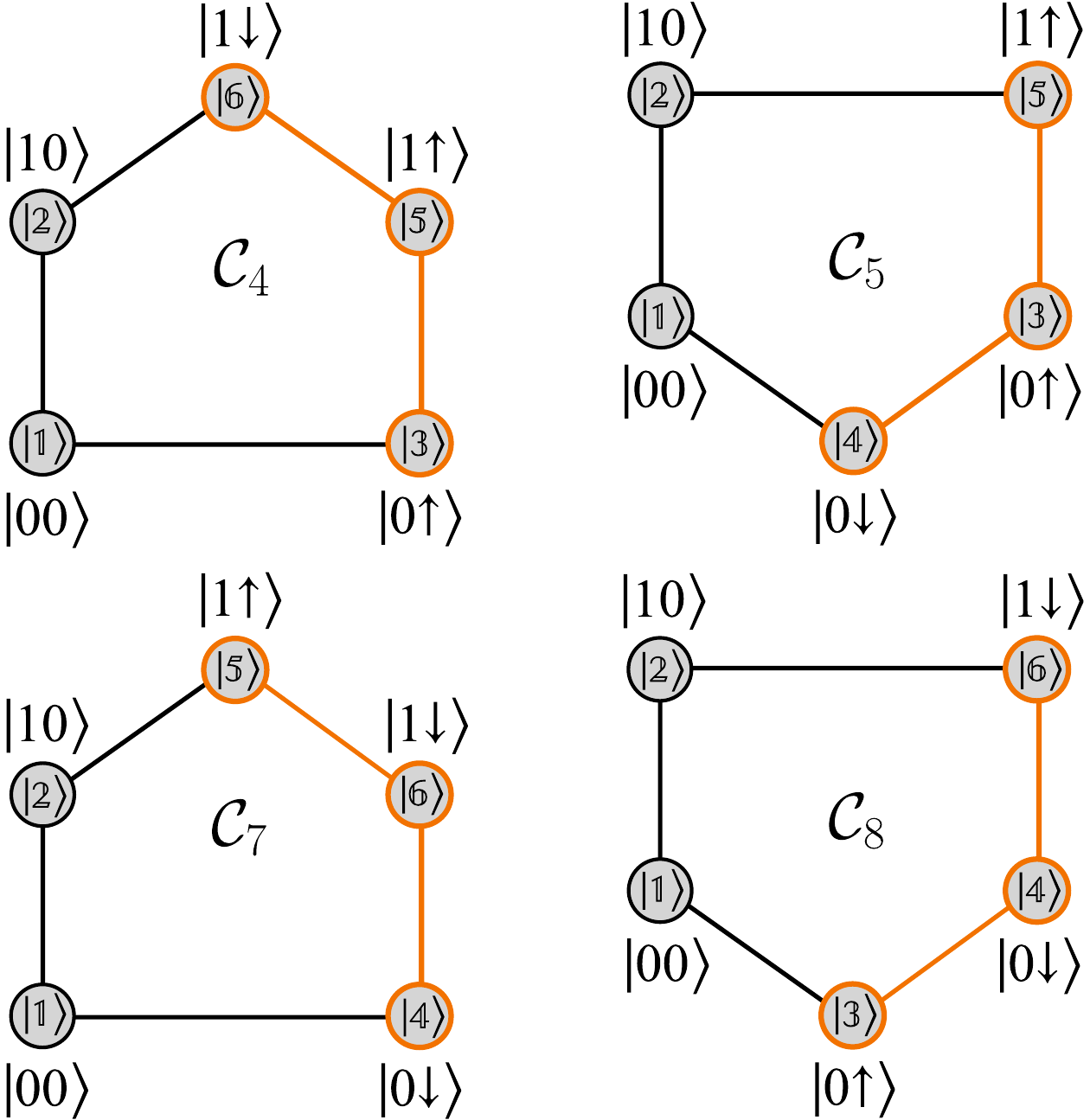}
    \caption{Schematic diagram of four subcycles $\{ \mathcal{C}_4,\mathcal{C}_5,\mathcal{C}_7,\mathcal{C}_8\}$ which are truly responsible for the spin-thermolectric cross-effect of SSE and SPE. Special edges are marked in orange.}
    \label{cycles-in-cross-effects}
\end{figure}

In Section~\ref{Onsager-relation}, we have identified cycles ${\mathcal{C}_1, \mathcal{C}_2, \mathcal{C}_{10}, \mathcal{C}_{11} }$, which can contribute solely to the spin current and not to energy current. Conversely, cycles ${\mathcal{C}_3, \mathcal{C}_6 }$ are found to contribute exclusively to the energy current and not to the spin current. Meanwhile, cycles ${\mathcal{C}_4, \mathcal{C}_5, \mathcal{C}_{7}, \mathcal{C}_{8} }$ have the quality to contribute to both energy as well as spin currents [Fig.~\ref{cycles-in-cross-effects}].  To gain a deeper understanding, it is essential to analyze the complete topology of the network. Notably, we observe that all cycles contributing to $J_{\rm S}$ must involve a spin-flip process, either $\ket{0\uparrow}\leftrightarrow\ket{0\downarrow}$ or $\ket{1\uparrow}\leftrightarrow\ket{1\downarrow}$, corresponding to the edges $\ket{\mathbb{3}}\leftrightarrow\ket{\mathbb{4}}$ or $\ket{\mathbb{5}}\leftrightarrow\ket{\mathbb{6}}$, respectively. Similarly, cycles contributing to $J_{\rm E}$ must include the edges $\ket{\mathbb{3}}\leftrightarrow\ket{\mathbb{5}}$ ($\ket{0\uparrow}\leftrightarrow\ket{1\uparrow}$) or $\ket{\mathbb{4}}\leftrightarrow\ket{\mathbb{6}}$ ($\ket{0\downarrow}\leftrightarrow\ket{1\downarrow}$), enabling Coulomb interaction between the upper and lower dots. This insight sheds light on why the expressions for spin and energy currents, derived from the master equation, take their particular forms [Cf.~Eqs.~\eqref{spin-current}~and~\eqref{energy-current}].
At this juncture, it's imperative to underscore that cycles ${\mathcal{C}_1, \mathcal{C}_2, \mathcal{C}_9, \mathcal{C}_{10}, \text{ and } \mathcal{C}_{11}}$ share the edge $\ket{\mathbb{3}}\leftrightarrow\ket{\mathbb{4}}$ or $\ket{\mathbb{5}}\leftrightarrow\ket{\mathbb{6}}$, yet they do not contribute to the spin current in the SSE due to zero cycle affinity. This results from the fact that the circuit fluxes associated with the cycle trajectories ($\mathcal{C}^+$ and $\mathcal{C}^-$) are identical in both directions, yielding a zero cycle flux. The same holds true for the SPE with cycles $\mathcal{C}_3$, $\mathcal{C}_6$, and $\mathcal{C}_9$ in the absence of a spin bias voltage, despite having the required edges. Intriguingly, cycle $\mathcal{C}_9$ possesses both spin-flip and Coulomb-interaction edges [Fig.~\ref{fig2}], yet it yields zero cycle fluxes due to its zero cycle affinity. Therefore, the asymmetry in cycle affinity emerges as the primary thermodynamic driving force and the foremost criterion for obtaining a nonzero cycle flux. Conversely, cycles ${ \mathcal{C}_4, \mathcal{C}_5, \mathcal{C}_7, \mathcal{C}_8}$ exhibit nonzero cycle affinity either in the absence of a temperature bias (SPE) or in the absence of a spin bias voltage (SSE). Consequently, these four cycles stand as the sole contributors to both SSE and SPE, featuring finite spin and energy currents. This is attributed to their possession of both the spin-flip edge ($\ket{\mathbb{3}}\leftrightarrow\ket{\mathbb{4}}$ or $\ket{\mathbb{5}}\leftrightarrow\ket{\mathbb{6}}$) and the Coulomb-interaction edge ($\ket{\mathbb{3}}\leftrightarrow\ket{\mathbb{5}}$ or $\ket{\mathbb{4}}\leftrightarrow\ket{\mathbb{6}}$). To summarize, we affirm that ${ \mathcal{C}_4, \mathcal{C}_5, \mathcal{C}_7, \mathcal{C}_8}$ stand as the four pivotal cycles solely responsible for materializing the thermodynamic cross-effect in the form of the spin-thermoelectric effect within our simple minimal model of the quantum thermocouple. The consequences of the interference effect between these major contributing cycles open intriguing avenues, and future research directions could delve into the impact of quantum coherence and entanglement effect~\cite{rao2020spin,whitney2016quantum} on the device performance from the perspective of the network theory.

\section{Conclusions}\label{conclusion}

The key findings of our present analysis are outlined as follows: (i) We present a simple model of a quantum thermocouple that exhibits spin-caloritronic effects based on three-terminal ultra-strong Coulomb-coupled quantum dots. In contrast to four-terminal models, this minimal model mimics both spin-dependent Seebeck and Peltier effects in complete analogy to classical thermocouples, used to describe thermoelectric effects. In the quantum case, distinct statistical properties of the thermal reservoirs play a role akin to dissimilar metals in traditional thermocouples. (ii) We find out that the expressions for spin and energy currents, derived from the Lindblad master equation, completely agree with network theoretical results. However, the quantum kinetic Pauli master equation serves as the basis for constructing the thermodynamic network, encompassing joint system microstates and associated transition rates. This is in stark contrast to classical network theory, where microstates often result from coarse-graining procedures. Here instead, they naturally emerge as eigenstates of the coupled quantum systems, derived from the microscopic Hamiltonian description of the composite quantum system. (iii) Benefiting from the generalized matrix tree theorem of the algebraic graph, we not only unveil the fundamental operational principles behind spin-Seebeck and spin-Peltier effects but also confirm the applicability of well-known thermodynamic relations in nano-thermoelectric devices. The validity of Onsager reciprocity and Kelvin relations for thermoelectric coefficients underscore the universal generality of thermodynamic principles in both classical and quantum realms.  In the present case, the above relations stem from the characteristic properties of forward and backward cycle flux trajectories of the quantum thermodynamic network. This is fundamentally different from the phenomenological classical laws of irreversible thermodynamics that hinge on local equilibrium assumptions. (iv) In this context, we stress the importance of network cycle flux, and cycle affinity in establishing the macroscopic spin and energy currents in terms of stochastic cycle currents. Cycle affinity, expressed as a ratio of transition rates between forward and backward cycle trajectories, emerges as a fundamental driving force behind nonzero cycle fluxes. Then, the cycle flux ranking scheme powered by the microscopic or stochastic version of the entropy production rate, sheds light on the origin of weak spin and energy currents in spin-Seebeck and Peltier effects, respectively. (v) Finally, we identify four non-intersecting cycles that are responsible for manifesting both reciprocal effects of spin-thermoelectricity within our simple minimal model. Characterized by special edges involving the spin-flip process and Coulomb interaction between interacting quantum dots, these cycles pave the way for underpinning the fundamental working principles of quantum thermocouples.

\section*{Acknowledgements}
We thank Sujan Kundu for the useful discussions. AG acknowledges financial support from the Initiation grant of IITK (Grant No. IITK/CHM/2018513). N.G. is thankful to CSIR for the fellowship. S.G. acknowledges the Ministry of Education, Government of India, for the Prime Minister Research Fellowship (PMRF).

\onecolumngrid 
\appendix

\section{Derivation of the Lindblad Quantum Master Equation}\label{Appendix-A}
The total Hamiltonian of the overall three-terminal setup is given by 
\begin{equation}
    H=H_{\rm S}+H_{\rm B}+H_{\rm I},
\end{equation}
where $H_{\rm S}$, $H_{\rm B}$, and $H_{\rm I}$ are the total Hamiltonian of the system, bath, and system-bath interaction, respectively. The interaction Hamiltonian $H_{\rm I}$ is defined as $H_{\rm I}=H_{\rm IL}+H_{\rm IM}+H_{\rm IR}$, wherein $H_{\rm IL(R)}$ denotes the interaction of the lower quantum dot (${\rm QD}_l$) with the left (right) bath, and $H_{\rm IM}$ represents the interaction of the upper quantum dot (${\rm QD}_u$) with the middle bath. The interaction Hamiltonian $H_{\rm I\alpha}$ ($\alpha={\rm L, M, R}$) for each $\alpha$-th bath is given by~\cite{wang2022cycleflux,gupt2023topranked}
\begin{eqnarray}\label{int-H-L-M-R}
    H_{\rm IL}&=&H_{\rm IL\uparrow}+H_{\rm IL\downarrow},\; H_{\rm IL\uparrow}=\hbar\sum_{k}(t_{{\rm L}k}b^\dagger_{{\rm L}\uparrow k} d_{l\uparrow}+t^*_{{\rm L}k} d^\dagger_{l\uparrow}b_{{\rm L}\uparrow k}),\; H_{\rm IL\downarrow}=\hbar\sum_{k}(t_{{\rm L}k}b^\dagger_{{\rm L}\downarrow k} d_{l\downarrow}+t^*_{{\rm L}k} d^\dagger_{l\downarrow}b_{{\rm L}\downarrow k}),\nonumber\\
    H_{\rm IM}&=& \hbar\sum_{k}(t_{{\rm M}k}b^\dagger_{{\rm M} k} d_{u}+t^*_{{\rm M}k} d^\dagger_{u}b_{{\rm M} k}),\;
    H_{\rm IR}= \hbar\sum_{q}(g_{{\rm R}q}a^\dagger_{{\rm R} q} d^\dagger_{l\uparrow} d_{l\downarrow}+g^*_{{\rm R}q} d^\dagger_{l\downarrow}d_{l\uparrow}a_{{\rm R} q}).
\end{eqnarray}
To formulate the master equation, we begin with the derivation by considering the von Neumann equation applied to the total density matrix $\rho_{\rm T}$ of the combined system and reservoirs in the interaction picture, as given in~\cite{breuer2002book}
\begin{equation}\label{von-Neumann-Eqn}
    \frac{d\rho_{\rm T}}{dt}=-\frac{i}{\hbar}[H_{\rm I}(t),\rho_{\rm T}(t)].
\end{equation}
Integrating Eq.\eqref{von-Neumann-Eqn} and tracing out the bath degrees of freedom, the master equation in terms of the reduced density matrix $\rho$ of the coupled quantum dot system under the Born-Markov approximation can be written as~\cite{breuer2002book}
\begin{equation}\label{masterEq}
    \frac{d\rho(t)}{dt}=-\frac{1}{\hbar^2}\Tr_B\int^\infty_0 ds [H_{\rm I}(t),[H_{\rm I}(t-s),\rho_{\rm T}(t)]].
\end{equation}
In Eq.\eqref{masterEq}, $\rho(t)=\Tr_{\rm B}\{\rho_{\rm T}\}\equiv\Tr_{\rm B}\{\rho(t)\otimes\rho_{\rm B}\}$ where $\rho_{\rm B}=\rho_{\rm L\uparrow}\otimes\rho_{\rm L\downarrow}\otimes\rho_{\rm M}\otimes\rho_{\rm R}$, and $\Tr_{\rm B}\equiv\Tr_{{\rm L\uparrow},{\rm L\downarrow},{\rm M},{\rm R}}$ stands for the trace over each bath degrees of freedom. As a result, we can rewrite Eq.~\eqref{masterEq} as~\cite{gupt2022PRE,shuvadip2022univarsal,gupt2023topranked}
\begin{equation}\label{ME_rho}
   \frac{d\rho(t)}{dt}=-\frac{1}{\hbar^2}\Tr_{{\rm L\uparrow},{\rm L\downarrow},{\rm M},{\rm R}}\int^\infty_0 ds [H_{\rm I}(t),[H_{\rm I}(t-s),\rho(t)\otimes\rho_{\rm L\uparrow}\otimes\rho_{\rm L\downarrow}\otimes\rho_{\rm M}\otimes\rho_{\rm R}]]. 
\end{equation}
Using the following relations~\cite{shuvadip2022univarsal,gupt2023topranked}
\begin{eqnarray}
   \Tr_{\rm L\uparrow}(b_{{\rm L}\uparrow k}(t) \rho_{\rm L\uparrow})&=&0= \Tr_{\rm L\uparrow}(b^\dagger_{{\rm L}\uparrow k}(t) \rho_{\rm L\uparrow}), \quad\quad \Tr_{\rm L\downarrow}(b_{{\rm L}\downarrow k}(t) \rho_{\rm L\downarrow})=0= \Tr_{\rm L\downarrow}(b^\dagger_{{\rm L}\downarrow k}(t) \rho_{\rm L\downarrow}) \\
   \Tr_{\rm M}(b_{{\rm M}k}(t) \rho_{\rm M})&=&0= \Tr_{\rm M}(b^\dagger_{{\rm M}k}(t) \rho_{\rm M}),\;\quad\quad
   \Tr_{\rm R}(a_{{\rm R}q}(t) \rho_{\rm R})=0= \Tr_{\rm R}(a^\dagger_{{\rm R}q}(t) \rho_{\rm R}), 
\end{eqnarray}
one can simplify Eq.\eqref{ME_rho} as~\cite{gupt2022PRE,shuvadip2022univarsal,gupt2023topranked}
\begin{equation}\label{simplied-MEq}
    \frac{d\rho(t)}{dt}=-\frac{1}{\hbar^2}\sum_\beta\Tr_{{\rm L\uparrow},{\rm L\downarrow},{\rm M},{\rm R}}\int^\infty_0 ds [H_{\rm I\beta}(t),[H_{\rm I\beta}(t-s),\rho(t)\otimes\rho_{\rm L\uparrow}\otimes\rho_{\rm L\downarrow}\otimes\rho_{\rm M}\otimes\rho_{\rm R}]],\quad \beta={\rm L\uparrow,L\downarrow,M,R}.
\end{equation}
Now, we use system operators in the interaction picture as
\begin{eqnarray}
    d_{\rm i}(t)&=&e^{{iH_{\rm S} t}/{\hbar}}d_{\rm i} e^{{-iH_{\rm S} t}/{\hbar}}=\sum_{\{\varepsilon_{\mathbb{ji}}\}} e^{{-i{\varepsilon_{\mathbb{ji}}} t}/{\hbar}}d_{\rm i}, \nonumber\\
    d^\dagger_{\rm i}(t)&=&e^{{iH_{\rm S} t}/{\hbar}}d^\dagger_{\rm i} e^{{-iH_{\rm S} t}/{\hbar}}=\sum_{\{\varepsilon_{\mathbb{ji}}\}} e^{{i{\varepsilon_{\mathbb{ji}}} t}/{\hbar}}d^\dagger_{\rm i},\quad  {\rm i}=l\uparrow,l\downarrow,u 
\end{eqnarray}
where $\varepsilon_\mathbb{ji}=\varepsilon_\mathbb{j}-\varepsilon_\mathbb{i}>0$ is the energy required for the transition between state $|\mathbb{i}\rangle$ and $|\mathbb{j}\rangle$ driven by their respective bath. Similarly, one can write the expressions for the bath operators in the interaction picture. With all these given prescriptions, we have simplified the Eq.\eqref{simplied-MEq}, resulting in the Lindblad form of the quantum master equation as follows:
\begin{equation}\label{Lindblad_MEqn}
  \frac{d\rho}{dt}=\mathcal{L}_{\rm L\uparrow}[\rho]+\mathcal{L}_{\rm L\downarrow}[\rho]+\mathcal{L}_{\rm M}[\rho]+\mathcal{L}_{\rm R}[\rho],
\end{equation}
The explicit forms of the Lindblad super operator $\mathcal{L}$ in the above equation are given by
\begin{eqnarray}\label{Lindblad_L-upspin}
   \mathcal{L}_{\rm L\uparrow}[\rho]=\sum_{\{\varepsilon_{\rm L\uparrow}\}} \gamma_{\rm L}\Big[ f(\varepsilon_{\rm L\uparrow},\mu_{\rm L\uparrow},T_{\rm L})\Big(d^\dagger_{l\uparrow}(\varepsilon_{\rm L\uparrow})\rho d_{l\uparrow}(\varepsilon_{\rm L\uparrow})-\frac{1}{2}\{d_{l\uparrow}(\varepsilon_{\rm L\uparrow})d^\dagger_{l\uparrow}(\varepsilon_{\rm L\uparrow}),\rho\}\Big) \nonumber\\
   + (1-f(\varepsilon_{\rm L\uparrow},\mu_{\rm L\uparrow},T_{\rm L}))\Big(d_{l\uparrow}(\varepsilon_{\rm L\uparrow})\rho d^\dagger_{l\uparrow}(\varepsilon_{\rm L\uparrow})-\frac{1}{2}\{d^\dagger_{l\uparrow}(\varepsilon_{\rm L\uparrow})d_{l\uparrow}(\varepsilon_{\rm L\uparrow}),\rho\}\Big) \Big],
\end{eqnarray}
\begin{eqnarray}\label{Lindblad_L-downspin}
   \mathcal{L}_{\rm L\downarrow}[\rho]=\sum_{\{\varepsilon_{\rm L\downarrow}\}} \gamma_{\rm L}\Big[ f(\varepsilon_{\rm L\downarrow},\mu_{\rm L\downarrow},T_{\rm L})\Big(d^\dagger_{l\downarrow}(\varepsilon_{\rm L\downarrow})\rho d_{l\downarrow}(\varepsilon_{\rm L\downarrow})-\frac{1}{2}\{d_{l\downarrow}(\varepsilon_{\rm L\downarrow})d^\dagger_{l\downarrow}(\varepsilon_{\rm L\downarrow}),\rho\}\Big) \nonumber\\
   + (1-f(\varepsilon_{\rm L\downarrow},\mu_{\rm L\downarrow},T_{\rm L}))\Big(d_{l\downarrow}(\varepsilon_{\rm L\downarrow})\rho d^\dagger_{l\downarrow}(\varepsilon_{\rm L\downarrow})-\frac{1}{2}\{d^\dagger_{l\downarrow}(\varepsilon_{\rm L\downarrow})d_{l\downarrow}(\varepsilon_{\rm L\downarrow}),\rho\}\Big) \Big],
\end{eqnarray}
\begin{eqnarray}\label{Lindblad_M}
   \mathcal{L}_{\rm M}[\rho]=\sum_{\{\varepsilon_{\rm M}\}} \gamma_{\rm M}\Big[ f(\varepsilon_{\rm M},\mu_{\rm M},T_{\rm L})\Big(d^\dagger_{u}(\varepsilon_{\rm M})\rho d_{u}(\varepsilon_{\rm M})-\frac{1}{2}\{d_{u}(\varepsilon_{\rm M})d^\dagger_{u}(\varepsilon_{\rm M}),\rho\}\Big) \nonumber\\
   + (1-f(\varepsilon_{\rm M},\mu_{\rm M},T_{\rm M}))\Big(d_{u}(\varepsilon_{\rm M})\rho d^\dagger_{u}(\varepsilon_{\rm M})-\frac{1}{2}\{d^\dagger_{u}(\varepsilon_{\rm M})d_{u}(\varepsilon_{\rm M}),\rho\}\Big) \Big],
\end{eqnarray}
\begin{eqnarray}\label{Lindblad_R}
   \mathcal{L}_{\rm R}[\rho]=\sum_{\{\varepsilon_{\rm R}\}} \gamma_{\rm R}\Big[ n(\varepsilon_{\rm R},T_{\rm R})\Big(V^\dagger_{l}(\varepsilon_{\rm R})\rho V_{l}(\varepsilon_{\rm R})-\frac{1}{2}\{V_{l}(\varepsilon_{\rm R})V^\dagger_{l}(\varepsilon_{\rm R}),\rho\}\Big) \nonumber\\
   + (n(\varepsilon_{\rm R},T_{\rm R})+1)\Big(V_{l}(\varepsilon_{\rm R})\rho V^\dagger_{l}(\varepsilon_{\rm R})-\frac{1}{2}\{V^\dagger_{l}(\varepsilon_{\rm R})V_{ l}(\varepsilon_{\rm R}),\rho\}\Big) \Big],
\end{eqnarray}
where the operators $V_l=d^\dagger_{l\uparrow}d_{l\downarrow}$ and $V^\dagger_l=d^\dagger_{l\downarrow}d_{l\uparrow}$ are responsible for the transition between spin-up ($\uparrow$) and spin-down ($\downarrow$) states. The transition rates corresponding to their respective bath are characterized by the various $\gamma$'s. The explicit form of all $\gamma$'s in terms of system-bath coupling constants can be calculated by Fermi's golden rule, as $\gamma_{\rm L}=2\pi\hbar\sum_{k}|t_{{\rm L}k}|^2 \delta\big(\varepsilon-\epsilon_{{\rm L\sigma} k}\big)$, where $\sigma=\{\uparrow,\downarrow\}$, $\gamma_{\rm M}=2\pi\hbar \sum_k  |t_{{\rm M}k}|^2 \delta\big(\varepsilon-\epsilon_{{\rm M}k}\big)$, and $\gamma_{\rm R}=2\pi\hbar \sum_q  |g_{{\rm R}q}|^2 \delta\big(\varepsilon-\epsilon_{{\rm R}q}\big)$~\cite{gupt2023topranked}. The functions $f(\varepsilon,\mu, T)=[e^{(\varepsilon-\mu)/k_B T}+1]^{-1}$ is the Fermi-Dirac distribution function corresponding to the left (L) and middle (M) bath with the transition energy $\varepsilon$, chemical potential $\mu$, and equilibrium bath temperature $T$. Similarly, the function $n(\varepsilon, T)=[e^{\varepsilon/k_B T}-1]^{-1}$ is the Bose-Einstein distribution function corresponding to the right (R) bath with the transition energy $\varepsilon$ and reservoir temperature $T$. The distribution functions are defined as the bath correlation functions and can be calculated as $\langle{b^\dagger b}\rangle=\Tr_{\rm L\sigma(M)}({b^\dagger b\rho_{\rm L\sigma(M)}})=f_{\rm L\sigma(M)}$ and $\langle{b b^\dagger}\rangle=\Tr_{\rm L\sigma(M)}({b b^\dagger\rho_{\rm L\sigma(M)}})=1-f_{\rm L\sigma(M)}$ for the left (middle) bath and $\langle{a^\dagger a}\rangle=\Tr_{\rm R}({a^\dagger a\rho_{\rm R}})=n_{\rm R}$, $\langle{a a^\dagger}\rangle=\Tr_{\rm R}({a a^\dagger\rho_{\rm R}})=1+n_{\rm R}$ for the right bath~\cite{gupt2023topranked}. The operators $b$ and $b^\dagger$ follow anti-commutation relation whereas the operators $a$ and $a^\dagger$ follow commutation relation, and $k_B$ is the Boltzmann constant. The energies needed for the transitions which are driven by the left and middle baths are $\varepsilon_{\rm L\uparrow}=\{\varepsilon_\mathbb{31},\varepsilon_\mathbb{52}\}$, $\varepsilon_{\rm L\downarrow}=\{\varepsilon_\mathbb{41},\varepsilon_\mathbb{62}\}$, and $\varepsilon_{\rm M}=\{\varepsilon_\mathbb{21},\varepsilon_\mathbb{53},\varepsilon_\mathbb{64}\}$ respectively, while the energies required for the transitions triggered by the right bath are $\varepsilon_{\rm R}=\{\varepsilon_\mathbb{43},\varepsilon_\mathbb{65}\}$. Note that one can express the various system creation and annihilation operators and their combinations in the following forms $|\mathbb{i}\rangle\langle\mathbb{j}|$ ($\mathbb{i}\ne\mathbb{j}$, $\mathbb{i},\;\mathbb{j}=\mathbb{1,2,3,4,5,6}$), which are given by:
\begin{eqnarray}\label{system-operators} d^\dagger_{l\uparrow}&=&|\mathbb{3}\rangle\langle\mathbb{1}|+|\mathbb{5}\rangle\langle\mathbb{2}|,\quad d_{l\uparrow}=|\mathbb{1}\rangle\langle\mathbb{3}|+|\mathbb{2}\rangle\langle\mathbb{5}|,\nonumber\\
d^\dagger_{l\downarrow}&=&|\mathbb{4}\rangle\langle\mathbb{1}|+|\mathbb{6}\rangle\langle\mathbb{2}|,\quad d_{l\downarrow}=|\mathbb{1}\rangle\langle\mathbb{4}|+|\mathbb{2}\rangle\langle\mathbb{6}|,
    \nonumber\\  d^\dagger_{u}&=&|\mathbb{2}\rangle\langle\mathbb{1}|+|\mathbb{5}\rangle\langle\mathbb{3}|+|\mathbb{6}\rangle\langle\mathbb{4}|,\quad d_{u}=|\mathbb{1}\rangle\langle\mathbb{2}|+|\mathbb{3}\rangle\langle\mathbb{5}|+|\mathbb{4}\rangle\langle\mathbb{6}|,\nonumber\\ V^\dagger_{l}&=&d^\dagger_{l\downarrow}d_{l\uparrow}=|\mathbb{4}\rangle\langle\mathbb{3}|+|\mathbb{6}\rangle\langle\mathbb{5}|,\quad V_{l}=d^\dagger_{l\uparrow}d_{l\downarrow}=|\mathbb{3}\rangle\langle\mathbb{4}|+|\mathbb{5}\rangle\langle\mathbb{6}|.
\end{eqnarray}
Finally, Eqs.~\eqref{P1-dot}~to~\eqref{P6-dot} in the main text can be derived with the help of Eqs.~\eqref{Lindblad_MEqn}-\eqref{system-operators} in the following manner. For example
\begin{equation}\label{dP/dt}
  \frac{dP_\mathbb{1}}{dt}=\langle\mathbb{1}|\frac{d\rho}{dt}|\mathbb{1}\rangle=\langle\mathbb{1}|\mathcal{L}_{\rm L\uparrow}[\rho]|\mathbb{1}\rangle+\langle\mathbb{1}|\mathcal{L}_{\rm L\downarrow}[\rho]|\mathbb{1}\rangle+\langle\mathbb{1}|\mathcal{L}_{\rm M}[\rho]|\mathbb{1}\rangle+\langle\mathbb{1}|\mathcal{L}_{\rm R}[\rho]|\mathbb{1}\rangle,
\end{equation}
where the terms
\begin{eqnarray}
    \langle\mathbb{1}|\mathcal{L}_{\rm L\uparrow}[\rho]|\mathbb{1}\rangle&=& {\gamma_{\rm L} f(\varepsilon_{\mathbb{31}},\mu_{\rm L\uparrow},T_{\rm L})}\Big(-\frac{1}{2}\langle \mathbb{1}|\mathbb{1}\rangle\langle \mathbb{1}|\rho|\mathbb{1}\rangle -\frac{1}{2} \langle \mathbb{1}|\rho|\mathbb{1}\rangle\langle \mathbb{1}|\mathbb{1}\rangle\Big) + \gamma_{\rm L} (1-f(\varepsilon_{\mathbb{31}},\mu_{\rm L\uparrow},T_{\rm L})) \Big(\langle \mathbb{1}|\mathbb{1}\rangle\langle \mathbb{3}|\rho|\mathbb{3}\rangle\langle \mathbb{1}|\mathbb{1}\rangle\Big)  \nonumber\\
   &=&\gamma_{\rm L} (1-f(\varepsilon_{\mathbb{31}},\mu_{\rm L\uparrow},T_{\rm L})) P_\mathbb{3} - \gamma_L f(\varepsilon_{\mathbb{31}},\mu_{\rm L\uparrow},T_{\rm L}) P_\mathbb{1} 
   = k_{\mathbb{13}}P_\mathbb{3}-k_{\mathbb{31}}P_\mathbb{1}\equiv J_\mathbb{13},\\
    \langle\mathbb{1}|\mathcal{L}_{\rm L\downarrow}[\rho]|\mathbb{1}\rangle&=& {\gamma_{\rm L} f(\varepsilon_{\mathbb{41}},\mu_{\rm L\downarrow},T_{\rm L})}\Big(-\frac{1}{2}\langle \mathbb{1}|\mathbb{1}\rangle\langle \mathbb{1}|\rho|\mathbb{1}\rangle -\frac{1}{2} \langle \mathbb{1}|\rho|\mathbb{1}\rangle\langle \mathbb{1}|\mathbb{1}\rangle\Big) + \gamma_{\rm L} (1-f(\varepsilon_{\mathbb{41}},\mu_{\rm L\downarrow},T_{\rm L})) \Big(\langle \mathbb{1}|\mathbb{1}\rangle\langle \mathbb{4}|\rho|\mathbb{4}\rangle\langle \mathbb{1}|\mathbb{1}\rangle\Big)  \nonumber\\
   &=&\gamma_{\rm L} (1-f(\varepsilon_{\mathbb{41}},\mu_{\rm L\downarrow},T_{\rm L})) P_\mathbb{4} - \gamma_L f(\varepsilon_{\mathbb{41}},\mu_{\rm L\downarrow},T_{\rm L}) P_\mathbb{1} 
   = k_{\mathbb{14}}P_\mathbb{4}-k_{\mathbb{41}}P_\mathbb{1}\equiv J_\mathbb{14},\\
  \langle\mathbb{1}|\mathcal{L}_{\rm M}[\rho]|\mathbb{1}\rangle&=& {\gamma_{\rm L} f(\varepsilon_{\mathbb{21}},\mu_{\rm L\downarrow},T_{\rm L})}\Big(-\frac{1}{2}\langle \mathbb{1}|\mathbb{1}\rangle\langle \mathbb{1}|\rho|\mathbb{1}\rangle -\frac{1}{2} \langle \mathbb{1}|\rho|\mathbb{1}\rangle\langle \mathbb{1}|\mathbb{1}\rangle\Big) + \gamma_{\rm L} (1-f(\varepsilon_{\mathbb{21}},\mu_{\rm L\downarrow},T_{\rm L})) \Big(\langle \mathbb{1}|\mathbb{1}\rangle\langle \mathbb{2}|\rho|\mathbb{2}\rangle\langle \mathbb{1}|\mathbb{1}\rangle\Big)  \nonumber\\
   &=&\gamma_{\rm L} (1-f(\varepsilon_{\mathbb{21}},\mu_{\rm L\downarrow},T_{\rm L})) P_\mathbb{2} - \gamma_L f(\varepsilon_{\mathbb{21}},\mu_{\rm L\downarrow},T_{\rm L}) P_\mathbb{1} 
   = k_{\mathbb{12}}P_\mathbb{2}-k_{\mathbb{21}}P_\mathbb{1}\equiv J_\mathbb{12},\\
   \langle\mathbb{1}|\mathcal{L}_{\rm R}[\rho]|\mathbb{1}\rangle&=& 0.
\end{eqnarray}
Similarly, one can derive time evolution equations for the population of the other $\mathbb{i}$-th states. Under the steady state, $dP_\mathbb{i}/dt=0$ ($\mathbb{i=1,2,..,6}$), and we have
\begin{eqnarray}
    \frac{dP_\mathbb{1}}{dt}&=&(k_{\mathbb{13}}P_\mathbb{3}-k_{\mathbb{31}}P_\mathbb{1})+(k_{\mathbb{14}}P_\mathbb{4}-k_{\mathbb{41}}P_\mathbb{1})+(k_{\mathbb{12}}P_\mathbb{2}-k_{\mathbb{21}}P_\mathbb{1})=J_\mathbb{13}+J_\mathbb{14}+J_\mathbb{12}=0,\\
    \frac{dP_\mathbb{2}}{dt}&=&(k_{\mathbb{25}}P_\mathbb{5}-k_{\mathbb{52}}P_\mathbb{2})+(k_{\mathbb{26}}P_\mathbb{6}-k_{\mathbb{62}}P_\mathbb{2})+(k_{\mathbb{21}}P_\mathbb{1}-k_{\mathbb{12}}P_\mathbb{2})=J_\mathbb{25}+J_\mathbb{26}+J_\mathbb{21}=0,\\
    \frac{dP_\mathbb{3}}{dt}&=&(k_{\mathbb{31}}P_\mathbb{1}-k_{\mathbb{13}}P_\mathbb{3})+(k_{\mathbb{35}}P_\mathbb{5}-k_{\mathbb{53}}P_\mathbb{3})+(k_{\mathbb{34}}P_\mathbb{4}-k_{\mathbb{43}}P_\mathbb{3})=J_\mathbb{31}+J_\mathbb{35}+J_\mathbb{34}=0,\\
    \frac{dP_\mathbb{4}}{dt}&=&(k_{\mathbb{41}}P_\mathbb{1}-k_{\mathbb{14}}P_\mathbb{4})+(k_{\mathbb{46}}P_\mathbb{6}-k_{\mathbb{64}}P_\mathbb{4})+(k_{\mathbb{43}}P_\mathbb{3}-k_{\mathbb{34}}P_\mathbb{4})=J_\mathbb{41}+J_\mathbb{46}+J_\mathbb{43}=0,\\
    \frac{dP_\mathbb{5}}{dt}&=&(k_{\mathbb{52}}P_\mathbb{2}-k_{\mathbb{25}}P_\mathbb{5})+(k_{\mathbb{53}}P_\mathbb{3}-k_{\mathbb{35}}P_\mathbb{5})+(k_{\mathbb{56}}P_\mathbb{6}-k_{\mathbb{65}}P_\mathbb{5})=J_\mathbb{52}+J_\mathbb{53}+J_\mathbb{56}=0,\\
    \frac{dP_\mathbb{6}}{dt}&=&(k_{\mathbb{62}}P_\mathbb{2}-k_{\mathbb{26}}P_\mathbb{6})+(k_{\mathbb{64}}P_\mathbb{4}-k_{\mathbb{46}}P_\mathbb{6})+(k_{\mathbb{65}}P_\mathbb{5}-k_{\mathbb{56}}P_\mathbb{6})=J_\mathbb{62}+J_\mathbb{64}+J_\mathbb{65}=0.
\end{eqnarray}
In the present case, there is no particle exchange between the quantum dots, implying no particle current due to the middle reservoir. So, the steady-state energy (heat)  current through the middle reservoir within the Born-Markov-Secular (BMS) master equation can be defined as~\cite{shuvadip2022univarsal,wang2022cycleflux}
\begin{equation}\label{heat-current-JM}
    J_{\rm E}=J^{\rm M}_{\rm E}=\Tr\{\mathcal{L}_{\rm M}[\rho]H_{\rm S}\},
\end{equation}
where the system Hamiltonian $H_{\rm S}$ has the following form, $H_{\rm S}=\sum_\mathbb{i}\varepsilon_\mathbb{i}|\mathbb{i}\rangle\langle\mathbb{i}|$, with $\varepsilon_\mathbb{i}$ being the energy of the $\mathbb{i}$-th state. Using Eq.\eqref{heat-current-JM}, the expression for t$J_{\rm E}$ can be calculated as
\begin{eqnarray}
    J_{\rm E}=\varepsilon_\mathbb{21} J_\mathbb{21}+\varepsilon_\mathbb{53} J_\mathbb{53}+\varepsilon_\mathbb{64} J_\mathbb{64}.\label{J_E}
\end{eqnarray}
At the steady state, one may verify $J_\mathbb{12}=-J_\mathbb{21}=J_\mathbb{53}+J_\mathbb{64}$. As a result, Eq.~\eqref{J_E} reduce to Eq.\eqref{energy-current} of the main text:
\begin{eqnarray}
    J_{\rm E}&=&\varepsilon_u (-J_\mathbb{12})+(\varepsilon_u + {\rm U}) J_\mathbb{53}+(\varepsilon_u + {\rm U}) J_\mathbb{64},\nonumber\\
    &=&-\varepsilon_u J_\mathbb{12}+(\varepsilon_u + {\rm U})(J_\mathbb{53}+J_\mathbb{64}),\nonumber\\
    &=&{\rm U}(J_\mathbb{53}+J_\mathbb{64}).
\end{eqnarray}
Similarly, the steady-state spin current due to the left and right reservoirs can be defined as~\cite{wang2022cycleflux}:
\begin{eqnarray}
    J^{\rm L}_{\rm S}&=&\frac{1}{2}\Big(\Tr\{d^\dagger_{l\downarrow}d_{l\downarrow}\mathcal{L}_{\rm L\downarrow}[\rho]\}-\Tr\{d^\dagger_{l\uparrow}d_{l\uparrow}\mathcal{L}_{\rm L\uparrow}[\rho]\}\Big),\label{spin-current-JL}\\
    J^{\rm R}_{\rm S}&=&\Tr\{V^\dagger_l V_l\mathcal{L}_{\rm R}[\rho]\}.\label{spin-current-JR}
\end{eqnarray}
Now, using Eq.\eqref{system-operators}, one can derive the expressions for $J^{\rm L}_{\rm S}$ and $J^{\rm R}_{\rm S}$ from Eqs.~\eqref{spin-current-JL} and \eqref{spin-current-JR} in the follwing forms:
\begin{eqnarray}
    J^{\rm L}_{\rm S}&=&\frac{1}{2}\Big[(J_\mathbb{41}+J_\mathbb{62})-(J_\mathbb{31}+J_\mathbb{52})\Big],\\
    J^{\rm R}_{\rm S}&=&J_\mathbb{43}+J_\mathbb{65}.
\end{eqnarray}
In the steady-state, $J_\mathbb{41}+J_\mathbb{62}=-(J_\mathbb{31}+J_\mathbb{52})=J_\mathbb{34}+J_\mathbb{56}$. As a result, we get Eq.\eqref{spin-current} of the main text as the steady-state spin current
\begin{eqnarray}
J_{\rm S}&=&J^{\rm L}_{\rm S}=\frac{1}{2}\Big[(J_\mathbb{41}+J_\mathbb{62})+(J_\mathbb{41}+J_\mathbb{62})\Big]\nonumber\\
&=&J_\mathbb{41}+J_\mathbb{62}=J_\mathbb{34}+J_\mathbb{56}\\
&=&-(J_\mathbb{43}+J_\mathbb{65})=-J^{\rm R}_{\rm S}.
\end{eqnarray}

\section{Derivation of the entropy production rate}\label{Appendix-B}

From Eq.\eqref{st-sigma}, the steady-state entropy production rate can be written as follows~\cite{schnakenberg1976network,landi2021irreversible}:
\begin{eqnarray}
    \dot{\sigma}&=&\frac{1}{2}k_B\sum_\mathbb{i,j} J_\mathbb{ij}\ln\Big(\frac{k_\mathbb{ij}}{k_\mathbb{ij}}\Big)\nonumber\\
    &=& \frac{1}{2}k_B\Bigg[ J_\mathbb{31}\ln\Big(\frac{k_\mathbb{31}}{k_\mathbb{13}}\Big)+J_\mathbb{13}\ln\Big(\frac{k_\mathbb{13}}{k_\mathbb{31}}\Big)+J_\mathbb{41}\ln\Big(\frac{k_\mathbb{41}}{k_\mathbb{14}}\Big)+J_\mathbb{14}\ln\Big(\frac{k_\mathbb{14}}{k_\mathbb{41}}\Big) +J_\mathbb{43}\ln\Big(\frac{k_\mathbb{43}}{k_\mathbb{34}}\Big) +J_\mathbb{34}\ln\Big(\frac{k_\mathbb{34}}{k_\mathbb{43}}\Big)+ ..... \Bigg]
\end{eqnarray}
As we have mentioned in the main text $J_\mathbb{ij}=-J_\mathbb{ji}$ for all $\mathbb{i}$ and $\mathbb{j}$ ($\mathbb{i\ne j}$), so $\dot{\sigma}$ will be equal to 
\begin{eqnarray}
    \dot{\sigma}&=&k_B\Bigg[ J_\mathbb{31}\ln\Big(\frac{k_\mathbb{31}}{k_\mathbb{13}}\Big)+J_\mathbb{41}\ln\Big(\frac{k_\mathbb{41}}{k_\mathbb{14}}\Big)+J_\mathbb{43}\ln\Big(\frac{k_\mathbb{43}}{k_\mathbb{34}}\Big)+J_\mathbb{21}\ln\Big(\frac{k_\mathbb{21}}{k_\mathbb{12}}\Big) +J_\mathbb{52}\ln\Big(\frac{k_\mathbb{52}}{k_\mathbb{25}}\Big)+J_\mathbb{64}\ln\Big(\frac{k_\mathbb{64}}{k_\mathbb{46}}\Big) + ..... \Bigg]\nonumber\\
    &=& k_B\Bigg[ (J^+_{\mathcal{C}_1}-J^-_{\mathcal{C}_1}+J^+_{\mathcal{C}_3}-J^-_{\mathcal{C}_3}+J^+_{\mathcal{C}_4}-J^-_{\mathcal{C}_4}+J^+_{\mathcal{C}_8}-J^-_{\mathcal{C}_8}+J^+_{\mathcal{C}_{11}}-J^-_{\mathcal{C}_{11}})\ln\Big(\frac{k_\mathbb{31}}{k_\mathbb{13}}\Big) + (-J^+_{\mathcal{C}_1}+J^-_{\mathcal{C}_1}+J^+_{\mathcal{C}_5}-J^-_{\mathcal{C}_5}+J^+_{\mathcal{C}_6}-J^-_{\mathcal{C}_6} \nonumber\\ 
    &+&J^+_{\mathcal{C}_7}-J^-_{\mathcal{C}_7}-J^+_{\mathcal{C}_{11}}+J^-_{\mathcal{C}_{11}})\ln\Big(\frac{k_\mathbb{41}}{k_\mathbb{14}}\Big) +(J^+_{\mathcal{C}_1}-J^-_{\mathcal{C}_1}-J^+_{\mathcal{C}_5}+J^-_{\mathcal{C}_5}+J^+_{\mathcal{C}_8}-J^-_{\mathcal{C}_8}-J^+_{\mathcal{C}_9}-J^-_{\mathcal{C}_9}-J^+_{\mathcal{C}_{10}}+J^-_{\mathcal{C}_{10}})\ln\Big(\frac{k_\mathbb{43}}{k_\mathbb{34}}\Big) 
    \nonumber\\
    &+& (-J^+_{\mathcal{C}_3}+J^-_{\mathcal{C}_3}-J^+_{\mathcal{C}_4}+J^-_{\mathcal{C}_4}-J^+_{\mathcal{C}_5}+J^-_{\mathcal{C}_5}-J^+_{\mathcal{C}_6}+J^-_{\mathcal{C}_6}-J^+_{\mathcal{C}_7}+J^-_{\mathcal{C}_7}-J^+_{\mathcal{C}_8}+J^-_{\mathcal{C}_8})\ln\Big(\frac{k_\mathbb{21}}{k_\mathbb{12}}\Big) +  .....  \Bigg]\nonumber\\
    &=& k_B\Bigg[ J^+_{\mathcal{C}_1}\ln\Bigg(\frac{k_\mathbb{14}k_\mathbb{43}k_\mathbb{31}}{k_\mathbb{13}k_\mathbb{34}k_\mathbb{41}} \Bigg)-J^-_{\mathcal{C}_1}\ln\Bigg(\frac{k_\mathbb{14}k_\mathbb{43}k_\mathbb{31}}{k_\mathbb{13}k_\mathbb{34}k_\mathbb{41}} \Bigg)+ ..... \Bigg] 
    = k_B\Bigg[ J^+_{\mathcal{C}_1}\ln\Bigg(\frac{\Pi^+_{\mathcal{C}_1}}{\Pi^-_{\mathcal{C}_1}} \Bigg)-J^-_{\mathcal{C}_1}\ln\Bigg(\frac{\Pi^+_{\mathcal{C}_1}}{\Pi^-_{\mathcal{C}_1}} \Bigg)+ ..... \Bigg]\nonumber\\
    &=& k_B (J^+_{\mathcal{C}_1} -J^-_{\mathcal{C}_1})\ln\Bigg(\frac{\Pi^+_{\mathcal{C}_1}}{\Pi^-_{\mathcal{C}_1}} \Bigg)+ .....\nonumber\\
    &=& k_B\sum_\mathcal{C} J_\mathcal{C} \mathcal{X}_\mathcal{C}, \quad \text{where} \quad J_\mathcal{C}= (J^+_{\mathcal{C}} -J^-_{\mathcal{C}}) \quad \text{and} \quad \mathcal{X}_\mathcal{C}=\ln\Bigg(\frac{\Pi^+_{\mathcal{C}}}{\Pi^-_{\mathcal{C}}}\Bigg).
    \label{app-sigma}
\end{eqnarray}
The ratio of $\Pi^\pm_\mathcal{C}$ is equal to the ratio of $J^\pm_\mathcal{C}$ for each cycle trajectory. These ratios are
\begin{eqnarray}
    \frac{J^+_{\mathcal{C}_1}}{J^-_{\mathcal{C}_1}}&=&\frac{\Pi^+_{\mathcal{C}_1}}{\Pi^-_{\mathcal{C}_1}}=\frac{k_\mathbb{43}k_\mathbb{31}k_\mathbb{14}}{k_\mathbb{13}k_\mathbb{34}k_\mathbb{41}}=e^{-{\Delta\mu_{\rm S}}/{k_B T_0}}\approx \Big(1-\frac{\Delta\mu_{\rm S}}{k_B T_0}\Big),
    \label{eqn-JC1}\\
    \frac{J^+_{\mathcal{C}_2}}{J^-_{\mathcal{C}_2}}&=&\frac{\Pi^+_{\mathcal{C}_2}}{\Pi^-_{\mathcal{C}_2}}=\frac{k_\mathbb{26}k_\mathbb{65}k_\mathbb{52}}{k_\mathbb{25}k_\mathbb{56}k_\mathbb{62}}
    =e^{-{\Delta\mu_{\rm S}}/{k_B T_0}}\approx \Big(1-\frac{\Delta\mu_{\rm S}}{k_B T_0}\Big),\\
    \frac{J^+_{\mathcal{C}_3}}{J^-_{\mathcal{C}_3}}&=&\frac{\Pi^+_{\mathcal{C}_3}}{\Pi^-_{\mathcal{C}_3}}=\frac{k_\mathbb{12}k_\mathbb{25}k_\mathbb{53}k_\mathbb{31}}{k_\mathbb{13}k_\mathbb{35}k_\mathbb{52}k_\mathbb{21}}
    =e^{{{\rm U}\delta T}/{k_B T_0(T_0 + \delta T)}}\approx \Big(1+\frac{{\rm U}\delta T}{k_B {T_0}^2}\Big),\\
    \frac{J^+_{\mathcal{C}_4}}{J^-_{\mathcal{C}_4}}&=&\frac{\Pi^+_{\mathcal{C}_4}}{\Pi^-_{\mathcal{C}_4}}=\frac{k_\mathbb{12}k_\mathbb{26}k_\mathbb{65}k_\mathbb{53}k_\mathbb{31}}{k_\mathbb{13}k_\mathbb{35}k_\mathbb{56}k_\mathbb{62}k_\mathbb{21}}
    =e^{{{\rm U}\delta T}/{k_B T_0(T_0 + \delta T)}}e^{-{\Delta\mu_{\rm S}}/{k_B T_0}}\approx \Big(1+\frac{{\rm U}\delta T}{k_B {T_0}^2}-\frac{\Delta\mu_{\rm S}}{k_B T_0}\Big),\\
    \frac{J^+_{\mathcal{C}_5}}{J^-_{\mathcal{C}_5}}&=&\frac{\Pi^+_{\mathcal{C}_5}}{\Pi^-_{\mathcal{C}_5}}=\frac{k_\mathbb{12}k_\mathbb{25}k_\mathbb{53}k_\mathbb{34}k_\mathbb{41}}{k_\mathbb{14}k_\mathbb{43}k_\mathbb{35}k_\mathbb{52}k_\mathbb{21}}=
    e^{{{\rm U}\delta T}/{k_B T_0(T_0 + \delta T)}}e^{{\Delta\mu_{\rm S}}/{k_B T_0}}\approx \Big(1+\frac{{\rm U}\delta T}{k_B {T_0}^2}+\frac{\Delta\mu_{\rm S}}{k_B T_0}\Big),\\
    \frac{J^+_{\mathcal{C}_6}}{J^-_{\mathcal{C}_6}}&=&\frac{\Pi^+_{\mathcal{C}_6}}{\Pi^-_{\mathcal{C}_6}}=\frac{k_\mathbb{12}k_\mathbb{26}k_\mathbb{64}k_\mathbb{41}}{k_\mathbb{14}k_\mathbb{46}k_\mathbb{62}k_\mathbb{21}}=
    e^{{{\rm U}\delta T}/{k_B T_0(T_0 + \delta T)}}\approx \Big(1+\frac{{\rm U}\delta T}{k_B {T_0}^2}\Big),\\
    \frac{J^+_{\mathcal{C}_7}}{J^-_{\mathcal{C}_7}}&=&\frac{\Pi^+_{\mathcal{C}_7}}{\Pi^-_{\mathcal{C}_7}}=\frac{k_\mathbb{12}k_\mathbb{25}k_\mathbb{56}k_\mathbb{64}k_\mathbb{41}}{k_\mathbb{14}k_\mathbb{46}k_\mathbb{65}k_\mathbb{52}k_\mathbb{21}}=
    e^{{{\rm U}\delta T}/{k_B T_0(T_0 + \delta T)}}e^{{\Delta\mu_{\rm S}}/{k_B T_0}}\approx \Big(1+\frac{{\rm U}\delta T}{k_B {T_0}^2}+\frac{\Delta\mu_{\rm S}}{k_B T_0}\Big),\\
    \frac{J^+_{\mathcal{C}_8}}{J^-_{\mathcal{C}_8}}&=&\frac{\Pi^+_{\mathcal{C}_8}}{\Pi^-_{\mathcal{C}_8}}=\frac{k_\mathbb{12}k_\mathbb{26}k_\mathbb{64}k_\mathbb{43}k_\mathbb{31}}{k_\mathbb{13}k_\mathbb{34}k_\mathbb{46}k_\mathbb{62}k_\mathbb{21}}=
    e^{{{\rm U}\delta T}/{k_B T_0(T_0 + \delta T)}}e^{-{\Delta\mu_{\rm S}}/{k_B T_0}}\approx \Big(1+\frac{{\rm U}\delta T}{k_B {T_0}^2}-\frac{\Delta\mu_{\rm S}}{k_B T_0}\Big),\\
    \frac{J^+_{\mathcal{C}_9}}{J^-_{\mathcal{C}_9}}&=&\frac{\Pi^+_{\mathcal{C}_9}}{\Pi^-_{\mathcal{C}_9}}= \frac{k_\mathbb{34}k_\mathbb{46}k_\mathbb{65}k_\mathbb{53}}{k_\mathbb{35}k_\mathbb{56}k_\mathbb{64}k_\mathbb{43}}=1, \\
    \frac{J^+_{\mathcal{C}_{10}}}{J^-_{\mathcal{C}_{10}}}&=&\frac{\Pi^+_{\mathcal{C}_{10}}}{\Pi^-_{\mathcal{C}_{10}}}=\frac{k_\mathbb{25}k_\mathbb{53}k_\mathbb{34}k_\mathbb{46}k_\mathbb{62}}{k_\mathbb{26}k_\mathbb{64}k_\mathbb{43}k_\mathbb{35}k_\mathbb{52}}=
    e^{{\Delta\mu_{\rm S}}/{k_B T_0}}\approx \Big(1+\frac{\Delta\mu_{\rm S}}{k_B T_0}\Big), \\
    \frac{J^+_{\mathcal{C}_{11}}}{J^-_{\mathcal{C}_{11}}}&=&\frac{\Pi^+_{\mathcal{C}_{11}}}{\Pi^-_{\mathcal{C}_{11}}}=\frac{k_\mathbb{14}k_\mathbb{46}k_\mathbb{65}k_\mathbb{53}k_\mathbb{31}}{k_\mathbb{13}k_\mathbb{35}k_\mathbb{56}k_\mathbb{64}k_\mathbb{41}}=
    e^{-{\Delta\mu_{\rm S}}/{k_B T_0}}\approx \Big(1-\frac{\Delta\mu_{\rm S}}{k_B T_0}\Big).
    \label{eqn-JC11}
\end{eqnarray}
Using Eqs.\eqref{eqn-JC1}-\eqref{eqn-JC11} into Eq.\eqref{app-sigma}, yields
\begin{eqnarray}
    \dot{\sigma}&=&J_{\mathcal{C}_1}\ln\Bigg( \frac{\Pi^+_{\mathcal{C}_{1}}}{\Pi^-_{\mathcal{C}_{1}}}\Bigg)+J_{\mathcal{C}_2}\ln\Bigg( \frac{\Pi^+_{\mathcal{C}_{2}}}{\Pi^-_{\mathcal{C}_{2}}}\Bigg)+J_{\mathcal{C}_3}\ln\Bigg( \frac{\Pi^+_{\mathcal{C}_{3}}}{\Pi^-_{\mathcal{C}_{3}}}\Bigg)+J_{\mathcal{C}_4}\ln\Bigg( \frac{\Pi^+_{\mathcal{C}_{4}}}{\Pi^-_{\mathcal{C}_{4}}}\Bigg)+J_{\mathcal{C}_5}\ln\Bigg( \frac{\Pi^+_{\mathcal{C}_{5}}}{\Pi^-_{\mathcal{C}_{5}}}\Bigg)+.....\nonumber\\
    &=& J_{\mathcal{C}_1}\Bigg(-\frac{\Delta\mu_{\rm S}}{k_B T_0} \Bigg)+J_{\mathcal{C}_2}\Bigg(-\frac{\Delta\mu_{\rm S}}{k_B T_0} \Bigg)+J_{\mathcal{C}_3}\Bigg(-\frac{{\rm U}\delta T}{k_B {T_0}(T_0+\delta T)} \Bigg)+J_{\mathcal{C}_4}\Bigg(\frac{{\rm U}\delta T}{k_B {T_0}(T_0+\delta T)} -\frac{\Delta\mu_{\rm S}}{k_B T_0}\Bigg)\nonumber\\
    &+& J_{\mathcal{C}_5}\Bigg(\frac{{\rm U}\delta T}{k_B {T_0}(T_0+\delta T)} +\frac{\Delta\mu_{\rm S}}{k_B T_0}\Bigg)+J_{\mathcal{C}_6}\Bigg(-\frac{{\rm U}\delta T}{k_B {T_0}(T_0+\delta T)}\Bigg)+.....\nonumber\\
    &=& {\rm U}[J_{\mathcal{C}_3}+J_{\mathcal{C}_4}+J_{\mathcal{C}_5}+J_{\mathcal{C}_6}+J_{\mathcal{C}_7}+J_{\mathcal{C}_8}]\frac{\delta T}{T_0(T_0+\delta T)} +[-J_{\mathcal{C}_1}-J_{\mathcal{C}_2}-J_{\mathcal{C}_4}+J_{\mathcal{C}_5}+J_{\mathcal{C}_7}-J_{\mathcal{C}_8}
    +J_{\mathcal{C}_{10}}-J_{\mathcal{C}_{11}}]\frac{\Delta\mu_{\rm S}}{T_0}\nonumber\\
    &=& J_{\rm E}\Big[\frac{1}{T_0}-\frac{1}{(T_0+\delta T)}\Big] + J_{\rm S}\Big[\frac{\mu_{\rm L\downarrow}}{T_0}-\frac{\mu_{\rm L\uparrow}}{T_0}\Big],
\end{eqnarray}
where we identify the expressions of the macroscopic energy and spin currents in terms of microscopic cycle fluxes 
\begin{eqnarray}
    J_{\rm E}&=&{\rm U}[J_{\mathcal{C}_3}+J_{\mathcal{C}_4}+J_{\mathcal{C}_5}+J_{\mathcal{C}_6}+J_{\mathcal{C}_7}+J_{\mathcal{C}_8}]\\
    J_{\rm S}&=& [-J_{\mathcal{C}_1}-J_{\mathcal{C}_2}-J_{\mathcal{C}_4}+J_{\mathcal{C}_5}+J_{\mathcal{C}_7}-J_{\mathcal{C}_8}
    +J_{\mathcal{C}_{10}}-J_{\mathcal{C}_{11}}],
\end{eqnarray}
which are equivalent to Eqs.~\eqref{EC-in-terms-of-cycle-flux} and~\eqref{SC-in-terms-of-cycle-flux} of the main text.

\end{document}